%% file: main.tex
\documentclass[8pt,a4paper,twocolumn]{article}
\usepackage[margin=20mm]{geometry}
\usepackage{cite}
\usepackage{amsmath,amssymb,amsfonts}
\usepackage{algorithmic}
\usepackage{graphicx}
\usepackage{textcomp}
\usepackage[colorlinks=true,linkcolor=blue,citecolor=blue,urlcolor=blue]{hyperref}

\usepackage{booktabs}
\usepackage{bm}
\newcommand\draft

\def\BibTeX{{\rm B\kern-.05em{\sc i\kern-.025em b}\kern-.08em
    T\kern-.1667em\lower.7ex\hbox{E}\kern-.125emX}}
\begin{document}

\input{src/00_title.tex}
\maketitle
\input{src/01_abst_e.tex}
\input{src/02_body_e.tex}
\input{src/03_append_e.tex}

\end{document}

%% file: src/00_title.tex
\title{Co-occurrence-Aware Quadratic Assignment for Local Feature Matching in Simultaneous Localization and Mapping}
\ifdefined\draft
\author{{Yutaka Yamada}$^{\ast}$,
{Yohei Hamakawa},
{Yutaro Ishigaki}, \\
{Masaya Yamasaki},
{Kosuke Tatsumura}
\\
\small Corporate Laboratory, Toshiba Corporation, Japan\\
$^{\ast}$\small Corresponding author: Yutaka Yamada (e-mail: yutaka.yamada.t68@mail.toshiba)
}
\date{}
\else
\author{\uppercase{Yutaka Yamada}\authorrefmark{1}, \IEEEmembership{Member, IEEE},
\uppercase{Yohei Hamakawa}\authorrefmark{1},
\uppercase{Yutaro Ishigaki}\authorrefmark{1}, \IEEEmembership{Member, IEEE},
\uppercase{Masaya Yamasaki}\authorrefmark{1}, and
\uppercase{Kosuke Tatsumura}\authorrefmark{1}, \IEEEmembership{Member, IEEE}}

\address[1]{Corporate Laboratory, Toshiba Corporation, Kawasaki 212-8582, Japan}

\markboth
{Y. Yamada \headeretal: Co-occurrence-Aware Quadratic Assignment for Local Feature Matching in Simultaneous Localization and Mapping}
{Y. Yamada \headeretal: Co-occurrence-Aware Quadratic Assignment for Local Feature Matching in Simultaneous Localization and Mapping}

\corresp{Corresponding author: Yutaka Yamada (e-mail: yutaka.yamada.t68@mail.toshiba).}
\fi

%% file: src/01_abst_e.tex
\begin{abstract}
Local feature matching, which associates keypoints in two images as keypoint pairs, is fundamental to Visual Simultaneous Localization and Mapping (Visual SLAM).
Nearest Neighbor (NN) search is commonly used for keypoint matching, but it has difficulty selecting correct keypoint pairs when multiple candidates have similar costs.
To improve matching accuracy, this paper proposes a keypoint matching method that considers the pairwise co-occurrence of two keypoint pairs.
The keypoint matching is formulated as a quadratic assignment problem, which is an NP-hard combinatorial optimization problem, making it difficult to solve quickly on conventional computers.
Recently, Ising machines have been developed as computing devices capable of solving hard combinatorial optimization problems.
Using a simulated bifurcation based Ising machine, the proposed method improved matching accuracy by approximately 8 percentage points over a conventional method on the HPatches dataset.
Furthermore, we integrated the proposed method into ORB-SLAM3, a representative academic Visual SLAM system, and achieved a 3.78-fold improvement in absolute pose error (APE) and a 2.85-fold improvement in relative pose error (RPE) on the KITTI dataset scenes where multiple same shape objects are repeatedly arranged, which are challenging for accurate self-pose estimation by the original ORB-SLAM3.
\end{abstract}

\ifdefined\draft
\relax
\else
\fi
 

%% file: src/02_body_e.tex
\section{Introduction} \label{sec:introduction}
\input{src/fig_yamad1.tex}

\ifdefined\draft
Local
\else
\PARstart{L}{ocal}
\fi
feature matching is a fundamental image processing method and is utilized in various applications such as Visual Simultaneous Localization and Mapping (Visual SLAM), Structure from Motion (SfM), and image recognition \cite{slam01_mur2015orb, slam01-2_mur2017orb, slam01-3_campos2021orb, snavely2006photo, klein2007parallel}.
Local feature matching consists of three processes: keypoint detection, feature extraction, and keypoint matching.
Keypoint detection detects keypoints from images and feature extraction extracts feature descriptors of detected keypoints.
Keypoint matching determines keypoint pairs, which represent the correspondence relationships between keypoints in the source and the reference images.

For keypoint detection and feature extraction, rule-based methods that manually design keypoint detection procedures and feature definitions have been widely proposed \cite{harris1988combined, rosten2006machine, lfm05_lowe2004distinctive, calonder2010brief, lfm04_rublee2011orb, lfm06_alcantarilla2011fast}.
In recent years, machine learning-based methods \cite{lfm07_detone2018superpoint, revaud2019r2d2, dusmanu2019d2} have also been proposed.

Keypoint matching selects corresponding keypoints from the detected keypoints in the reference image for the keypoints detected in the source image, and determines keypoint pairs.
Nearest Neighbor (NN) search selects the keypoint pair with the minimum cost and is widely used in keypoint matching.
Here, the cost represents the evaluation metric for searching corresponding keypoints between the source and the reference images.
In conventional keypoint matching, feature distance is used as the cost, where a smaller feature distance indicates a lower cost.
Machine learning-based methods \cite{lfm02_sarlin2020superglue, lfm01_sun2021loftr, lfm03_lindenberger2023lightglue} have also been devised for keypoint matching.
Machine learning-based keypoint matching commonly uses the likelihood of keypoint pairs as the cost, where a higher likelihood indicates a lower cost.
Machine learning-based methods achieve higher accuracy than conventional methods that use only feature distance by utilizing additional information such as position information for likelihood computation.

Since NN search simply selects the keypoint pair with the minimum cost, keypoint matching based on NN search may select a suboptimal pair due to cost evaluation errors when multiple keypoint pairs with similar costs exist.
For instance, when multiple objects of the same shape exist in both the source and reference images, a keypoint pair within the margin of error but with slightly lower cost may be selected, resulting in the matched keypoint pair not being the truly correct one. This causes problems such as correspondence swaps between keypoints of identically shaped objects, preventing correct correspondences.

Visual SLAM simultaneously performs camera pose estimation and surrounding environment map estimation from images captured by a camera.
Visual SLAM is a major application that uses local feature matching and is a fundamental technology for autonomous driving vehicles.
One of the main approaches to Visual SLAM is feature-based Visual SLAM, such as ORB-SLAM \cite{slam01_mur2015orb, slam01-2_mur2017orb, slam01-3_campos2021orb}.
Feature-based Visual SLAM performs pose estimation and map estimation using the results of local feature matching.
Therefore, the accuracy of local feature matching significantly affects the accuracy of feature-based Visual SLAM.
Correspondence swaps in keypoint pairs (erroneous matching) manifest as adverse effects such as incorrect camera-pose estimation in Visual SLAM.
In particular, when objects of the same shape such as guardrail poles are continuously arranged, incorrect keypoint pairs are likely selected.
Furthermore, when there are no other landmarks present, it becomes difficult to reliably exclude incorrect keypoint pairs as outliers, which complicates accurate pose estimation.
Such scenes are commonly encountered while driving and can be seen, for example, in Sequence 01 (a bridge scene) of the KITTI dataset \cite{Geiger2012CVPR} for SLAM and autonomous driving.
For autonomous driving vehicles, pose estimation errors can lead to misdetection of driving paths and potentially severe consequences, such as unexpected vehicle behavior.

To address such problems, this paper proposes ``Co-occurrence-aware Keypoint Matching,'' a keypoint matching method that selects more correct corresponding keypoint pairs using the ``pairwise co-occurrence likelihood of two keypoint pairs.''
The pairwise co-occurrence likelihood of two keypoint pairs is defined as the likelihood that two keypoint pairs are simultaneously correct correspondences.
When the camera pose and object motion between the source and reference images are known, ideally all corresponding keypoint pairs can be correctly generated.
In this case, the co-occurrence likelihood can similarly be correctly computed.
In practical cases, since the camera pose and object motion are unknown, the pairwise co-occurrence likelihood of two keypoint pairs is estimated by assuming the camera pose and object motion.
Using an example of performing local feature matching on two images of stationary objects captured by a horizontally moving camera, the pairwise co-occurrence likelihood of two keypoint pairs is explained.
In this example, the relative positional relationships of objects captured in the two images do not change significantly.
Therefore, the pairwise co-occurrence likelihood of two keypoint pairs is high for two keypoint pairs where the relative positional relationship of the keypoint pairs is maintained, and low for two keypoint pairs where the positional relationship is not maintained.
In addition to the feature cost related to feature distance and the geometry cost related to geometric distance used in conventional methods, Co-occurrence-aware Keypoint Matching introduces a new co-occurrence cost related to the pairwise co-occurrence likelihood of two keypoint pairs.
By considering the co-occurrence cost, correct corresponding keypoint pairs can be selected even when multiple keypoint pairs with similar features and geometric distances exist.

Co-occurrence-aware Keypoint Matching cannot be solved efficiently by NN search and requires formulation as a quadratic assignment problem (QAP).
The QAP is known as a representative NP-hard problem in combinatorial optimization \cite{cela2013quadratic}.
Such combinatorial optimization problems are difficult to solve optimally at high speed on conventional computers.
Ising machines are special-purpose computers designed to rapidly find optimal or near-optimal solutions to NP-hard combinatorial optimization problems.

Ising machines search for the minimum energy state in Ising spin models \cite{motcite44_brush1967history}.
The Ising problem is mathematically equivalent to quadratic unconstrained binary optimization (QUBO) and is known to be NP-hard \cite{motcite45_barahona1982computational}.
Various types of computationally difficult combinatorial optimization problems can be formulated as Ising problems \cite{ motcite46_lucas2014ising}.
Ising machines have been implemented on various hardware \cite{motcite17_finocchio2024roadmap}, including superconducting qubits \cite{motcite25_johnson2011quantum, motcite26_king2023quantum},
optical systems \cite{motcite29_bohm2019poor, motcite28_kalinin2020polaritonic, motcite27_honjo2021100},
oscillator-based circuits \cite{motcite18_goto2019combinatorial, motcite34_graber2024integrated, motcite35_moy20221, motcite36_albertsson2021ultrafast, motcite37_wang2021solving},
analog circuits \cite{motcite38_sharma2022increasing}, and digital circuits including ASIC \cite{motcite40_matsubara2020digital, motcite39_kawamura2023amorphica} and FPGA \cite{motcite18_goto2019combinatorial, motcite19_tatsumura2019fpga, motcite20_goto2021high, motcite21_tatsumura2021scaling, motcite23_kashimata2024efficient, motcite24_matsumoto2022distance, motcite60_hidaka2023correlation, tatsumura2025enhancing, motcite41_waidyasooriya2019highly, motcite42_leleu2021scaling}.
Quantum and optics-based implementations require special operating environments, limiting their availability to locations such as data centers.
On the other hand, digital circuit-based implementations can operate in environments equivalent to conventional computers and are expected to be deployable in embedded devices such as vehicles \cite{tatsumura2025enhancing}.
By using such Ising machines, local feature matching including QAPs can be solved in realistic time and with realistic implementation forms.
The scale of problems that an Ising machine can solve is limited by the number of spin variables it can handle and its connectivity network.
To utilize Ising machines more efficiently, methods that can match many keypoint pairs with fewer spin variables are important.

In this paper, we formulate Co-occurrence-aware Keypoint Matching as QUBO for processing on an Ising machine.
The objective function is composed of a linear combination of:
a feature function that minimizes the feature cost between keypoints of a single keypoint pair,
a geometry function that minimizes the geometry cost between keypoints of a single keypoint pair,
a co-occurrence function that minimizes the co-occurrence cost of two keypoint pairs, and
a penalty function corresponding to the constraint that a single keypoint can form a keypoint pair with at most one other keypoint.
We use an FPGA-based simulated bifurcation-machine \cite{motcite19_tatsumura2019fpga, motcite60_hidaka2023correlation} as an embeddable Ising machine to solve the QUBO.

We also demonstrate the effectiveness of the proposed method by integrating Co-occurrence-aware Keypoint Matching into local feature matching and feature-based Visual SLAM.
Figure \ref{fig:yamad1_concept} shows the feature-based Visual SLAM system with the proposed method.
This system follows conventional Visual SLAM processing, except that it uses Co-occurrence-aware Keypoint Matching, where keypoint pairs are determined by solving an optimization problem using an Ising machine.

The main contributions of this paper are as follows.
\begin{itemize}
\item Proposal of Co-occurrence-aware Keypoint Matching; \\
    We propose Co-occurrence-aware Keypoint Matching and formulate it as QUBO for solving on an Ising machine.
    Furthermore, since the scale of QUBO problems that an Ising machine can solve is constrained by the corresponding number of spin variables that the Ising machine has, we devised a method that enables keypoint matching for more keypoints with fewer bits.
\item Quantitative evaluation of the proposed keypoint matching in local feature matching; \\
    We conduct accuracy evaluation of local feature matching using the HPatches dataset \cite{hpatches_2017_cvpr}.
    The proposed method improves the accuracy of local feature matching by approximately 8 percentage points (\textbf{pp}) compared with the baseline.
\item Quantitative evaluation of the proposed keypoint matching in Visual SLAM; \\
    We conduct accuracy evaluation of pose estimation in Visual SLAM using the KITTI dataset \cite{Geiger2012CVPR}.
    In scenes where multiple objects of the same shape exist, the proposed method achieved 3.78-fold improvement in the Absolute Pose Error (APE) and 2.85-fold improvement in the Relative Pose Error (RPE) compared to the baseline.
\end{itemize}

The structure of this paper is as follows.
Section \ref{sec:relatedworks} describes related works and challenges in local feature matching and SLAM.
Section \ref{sec:lfmising} proposes a keypoint matching method using an Ising machine and also presents the local feature matching system and SLAM system incorporating the proposed method.
Section \ref{sec:evaluation} presents the evaluation of the proposed method.
Section \ref{sec:conclusion} presents the conclusion of this paper.

\section{Related Works} \label{sec:relatedworks}

\subsection{Local feature matching} \label{sec:localfeaturematching}

\input{src/fig_yamad2.tex}

Local feature matching finds correspondences between keypoints detected in the source image and those detected in the reference image, and generates keypoint pairs.
Figure \ref{fig:yamad2_lfm} shows the processing of conventional local feature matching.
Local feature matching consists of keypoint detection, feature extraction, and keypoint matching.

Keypoint detection (e.g.,  Harris-Corner and FAST \cite{harris1988combined, rosten2006machine}) detects keypoints that can be stably detected in images, such as edges and corners.
Feature extraction extracts feature descriptors (e.g., SIFT and BRIEF \cite{lfm05_lowe2004distinctive, calonder2010brief}) from detected keypoint neighborhoods.
Integrated methods such as ORB and AKAZE \cite{lfm04_rublee2011orb, lfm06_alcantarilla2011fast} combine both processes to improve quality.
In recent years, machine learning-based methods \cite{lfm07_detone2018superpoint, revaud2019r2d2, dusmanu2019d2} have been proposed.
While machine learning-based methods demonstrate high accuracy, they often require significant computational resources.

Keypoint matching searches for corresponding keypoints between the source and reference images and generates keypoint pairs according to the best matching metric.
NN search, which is widely used for keypoint matching, selects the keypoint pairs with the minimum cost, which is a metric of correspondence between source and reference image keypoints.
Conventional NN search commonly uses feature cost based on feature distance between keypoints of a keypoint pair \cite{lfm05_lowe2004distinctive, lfm04_rublee2011orb, lfm07_detone2018superpoint}.
The feature cost takes a smaller value as the feature distance decreases.
When the feature descriptor is a vector, the method for computing feature distance commonly uses Hamming distance or Euclidean distance.
Machine learning-based keypoint matching methods \cite{lfm02_sarlin2020superglue, lfm01_sun2021loftr, lfm03_lindenberger2023lightglue} have also been proposed and these methods compute the likelihood as cost from not only feature descriptors but also other image information such as position information \cite{lfm02_sarlin2020superglue, lfm03_lindenberger2023lightglue}.
The likelihood is treated as a monotonically decreasing function of cost, converting it so that a higher likelihood corresponds to a lower cost.
Although these machine learning-based methods achieve higher accuracy than conventional methods, they often require significant computational resources.

NN search selects the pair with the lowest cost.
Therefore, even if the cost of a correct keypoint pair is only slightly higher than that of an incorrect pair within the margin of error, the correct keypoint pair is not selected.
For example, if both the input and the reference image contain multiple objects of the same shape, the feature costs of keypoint pairs will be similar.
Furthermore, even when geometry cost is considered, if multiple identically shaped objects are placed nearby, the geometry costs also become similar.
In such cases, it becomes difficult to select the correct corresponding keypoint pairs based on feature cost and geometry cost.

\subsection{Visual Simultaneous Localization and Mapping} \label{sec:slam}

\input{src/fig_yamad3.tex}

Visual Simultaneous Localization and Mapping (SLAM) estimates the self-position and creates a 3D map of the surrounding area based on images captured by a camera (such as a monocular camera, a stereo camera, and an RGB-D camera).
Visual SLAM is a fundamental technology for autonomous mobility such as autonomous driving vehicles, AGVs, and drones.

Figure \ref{fig:yamad3_slam} shows the processing flow of Visual SLAM with a monocular camera configuration in ORB-SLAM and its derivatives \cite{slam01_mur2015orb, slam01-2_mur2017orb, slam01-3_campos2021orb}.
ORB-SLAM uses local feature matching and those features are ORB features \cite{lfm04_rublee2011orb}.
ORB-SLAM includes three processes: Tracking, Local Mapping, and Loop Closing.
The Tracking process performs local feature matching between the input frame and the previous frame from the camera to generate keypoint pairs, estimates the self-position, and tracks the keypoints.
Additionally, the Tracking process determines whether the input frame qualifies as a ``keyframe.''
The Mapping process generates map points of the 3D map using the self-position and the keypoints of the keyframe and optimizes the local map generated from the keyframe and neighboring frames (Local Bundle Adjustment).
The Loop Closing process detects loops on the global map from the self-position of keyframes, and optimizes the global map when a loop exists (Global Bundle Adjustment).
Since more correct keypoint pairs enable more accurate self-position estimation \cite{slam01_mur2015orb, slam01-2_mur2017orb, slam01-3_campos2021orb}, high-quality local feature matching is an important factor for Visual SLAM quality \cite{slam03_yurtsever2020survey, slam04_bresson2017simultaneous}.

As described in Section \ref{sec:introduction}, in scenes where objects of the same shape are continuously arranged, local feature matching using NN search might not select correct corresponding keypoint pairs, and correspondence swaps occur in keypoint pairs.
These problems lead to degradation of self-position estimation quality and map quality.
Furthermore, in scenes with few other landmarks, accurate self-pose estimation is more difficult.
If such scenes continue, the error in self-pose estimation will increase, which could ultimately lead to critical problems, such as unexpected behavior of an autonomous mobility.
In particular, these problems are prone to occur in open areas such as highways and bridges, or indoor environments such as tunnels and warehouses, where multiple objects of the same shape (such as guardrail poles, building windows, tunnel lighting, and shelves) are arranged with few other landmarks.
These scenes are frequently encountered during automobile driving, and solving problems that occur in such scenes is important.

\section{Co-occurrence-Aware Quadratic Assignment for Local Feature Matching} \label{sec:lfmising}

\input{src/tab_yamad1.tex}
\input{src/fig_yamad4.tex}

This section describes the proposed Co-occurrence-aware Keypoint Matching, which considers the pairwise co-occurrence likelihood of two keypoint pairs.
We formulate Co-occurrence-aware Keypoint Matching as an optimization problem on QUBO.
See Appendix \ref{sec:append-isingqubo} for the details of QUBO formulation.
This section also describes the local feature matching system and the Visual SLAM system incorporating the proposed method.
Table \ref{tab:yamad1_def} shows the definitions of variables used in this and subsequent sections.

As described in Section \ref{sec:localfeaturematching}, NN search, which is a common method in conventional local feature matching, may fail to select the correct keypoint pairs when there are multiple keypoint pair candidates with similar costs.
To select more correct corresponding keypoint pairs from multiple keypoint pair candidates with similar costs, we propose a new cost metric, ``co-occurrence cost'', which represents the pairwise co-occurrence likelihood of two keypoint pairs, in addition to the feature cost and geometry cost of keypoint pairs used in conventional methods.

When the self-pose and object motion between the source and reference images are known, ideally all corresponding keypoint pairs can be correctly generated.
Since all keypoint pair correspondences can be correctly generated, the pairwise co-occurrence likelihood of two keypoint pairs can similarly be correctly computed.
In many use cases, the self-pose and object motion are unknown.
Therefore, the pairwise co-occurrence likelihood of two keypoint pairs is estimated by assuming the self-pose and object motion.
We explain the assumptions about self-pose and object motion in monocular-camera Visual SLAM.
In Visual SLAM, the source and reference images use two consecutive frames, and the time interval between two consecutive frames is as small as 100 msec at 10 FPS.
Since the time interval between frames is short, the change in self-pose between frames is also small.
Objects captured by the camera are mostly stationary objects on background, and even for moving objects, the displacement between frames is small.
From these facts, it can be assumed that both the change in self-pose and the change in object motion between two frames in Visual SLAM are small.

Figure \ref{fig:yamad4_cooccurrence}(a)-(d) shows keypoints in a scene where objects of the same shape are continuously arranged.
In this figure, keypoints $s_0, s_1, s_2$ in the left source image correspond to keypoints $r_0, r_1, r_2$ in the right reference image, respectively.
In the left source image relative to the right reference image, keypoints have generally moved from right to left, and their relative positional relationships are maintained.
In this case, the positional relationships between two keypoint pairs with correct correspondences are also maintained.
Figure \ref{fig:yamad4_cooccurrence} (a) is an example of correct correspondences, where the positional relationships between each keypoint pair are maintained.
On the other hand, Figure \ref{fig:yamad4_cooccurrence} (b) is an example where keypoint pairs $(s_0, r_1)$ and $(s_1, r_0)$ are incorrect correspondences.
In this example, a swap of relative positional relationships occurs between keypoint pairs $(s_0, r_1)$ and $(s_1, r_0)$.
Similarly, Figure \ref{fig:yamad4_cooccurrence} (c) shows keypoint pairs $(s_0, r_2)$ and $(s_2, r_0)$ as incorrect correspondences,
and Figure \ref{fig:yamad4_cooccurrence} (d) shows keypoint pairs $(s_1, r_2)$ and $(s_2, r_1)$ as incorrect correspondences.
In these examples as well, swaps of relative positional relationships occur between keypoint pairs $(s_0, r_2)$ and $(s_2, r_0)$, and between keypoint pairs $(s_1, r_2)$ and $(s_2, r_1)$.
The pairwise co-occurrence likelihood of two keypoint pairs is set high when the positional relationship between the two keypoint pairs is maintained; and the likelihood is set low when the relationship is not maintained.
This makes it possible to select keypoint pairs that maintain relative positional relationships.
This leads to the possibility of selecting more correct corresponding keypoint pairs than when searching with only the feature cost and geometry cost.
Furthermore, it prevents relative positional swaps between keypoint pairs.

Figure \ref{fig:yamad4_cooccurrence}(e) shows the QUBO coefficient matrix (Q matrix) of the Co-occurrence-aware Keypoint Matching method presented in Section \ref{sec:lfmising-formulation}.
The size of this Q matrix is $(N_s \times N_r) \times (N_s \times N_r)$.
There are $N_s \times N_r$ decision variables (binary variables), each of which corresponds to a keypoint pair, $(s_i, r_j)$.
The diagonal terms of the Q matrix are linear coefficients corresponding to each of the $N_s \times N_r$ decision variables, and the off-diagonal terms are quadratic coefficients corresponding to pairs of two decision variables. The diagonal terms ($S_{i,j}$) include the feature cost and the geometry cost of one keypoint pair as used in conventional methods, and the off-diagonal terms include the co-occurrence cost of two keypoint pairs and constraint condition elements described in Section \ref{sec:lfmising-formulation}.
The gray areas in the off-diagonal terms indicate elements where keypoint pairs cannot be established due to constraint conditions.
While conventional methods search for keypoint pairs using only the diagonal terms of this Q matrix, the proposed method searches for keypoint pairs using both the diagonal terms and the off-diagonal terms.

Using the examples of Figure \ref{fig:yamad4_cooccurrence} (a), (b), (c), (d), we describe the assignment of co-occurrence likelihood in Figure \ref{fig:yamad4_cooccurrence} (e).
Although co-occurrence likelihood is originally represented by real numbers, Figure \ref{fig:yamad4_cooccurrence} (e) uses $++, +, -, --$ in descending order of likelihood for simplicity.
The two keypoint pairs $((s_0, r_0), (s_1, r_1))$ maintain their relative positional relationship.
Therefore, the elements $((s_0, r_0), (s_1, r_1))$ and $((s_1, r_1), (s_0, r_0))$ of this matrix have high co-occurrence likelihood ($++$).
Similarly, the matrix elements corresponding to two keypoint pairs $((s_0, r_0), (s_2, r_2))$ and $((s_1, r_1), (s_2, r_2))$ also have high co-occurrence likelihood ($++$).
On the other hand, the two keypoint pairs $((s_0, r_1), (s_1, r_0))$ do not maintain their relative positional relationship.
Therefore, the elements $((s_0, r_1), (s_1, r_0))$ and $((s_1, r_0), (s_0, r_1))$ of this matrix have low co-occurrence likelihood ($-$).
Similarly, the matrix elements corresponding to two keypoint pairs $((s_1, r_2), (s_2, r_1))$ also have low co-occurrence likelihood ($-$).
The matrix elements corresponding to two keypoint pairs $((s_0, r_2), (s_2, r_0))$ have even lower co-occurrence likelihood ($--$) due to the larger difference in relative positional relationships.

When simultaneously searching for the optimal combination of all pairs using a cost matrix containing co-occurrence likelihood elements,
it is impossible to use a simple method like NN search as a search method.
We formulate the keypoint pair search problem in Co-occurrence-aware Keypoint Matching as a QAP.
The QAP is known as one of the representative NP-hard problems in combinatorial optimization \cite{cela2013quadratic}.
Since it is difficult to quickly obtain optimal or near-optimal solutions for such combinatorial optimization problems on conventional computers, the proposed method realizes keypoint pair search by using an Ising machine.

\subsection{Formulation} \label{sec:lfmising-formulation}

This section formulates the problem of finding keypoint pairs in Co-occurrence-aware Keypoint Matching as an optimization problem on QUBO.
The optimization problem is to find the decision variables, $\mathbf{b}$, that minimize the objective function, $H_{\text{total}}(\mathbf{b})$.
The objective function $H_{\text{total}}(\mathbf{b})$ of the optimization problem for finding keypoint pairs is defined as follows.
\begin{align}
    H_{\text{total}}(\mathbf{b})
        &= \sum_{\substack{0 \le i < N_s \\ 0 \le j < N_r}} 
            \sum_{\substack{0 \le i' < N_s \\ 0 \le j' < N_r}}
                Q_{i,j,i',j'} \, b_{i,j} \, b_{i',j'}
\end{align}
$N_s$ and $N_r$ indicate the number of keypoints in the input image and the reference image, respectively.
The Q matrix, ${\mathbf{Q}} = (Q_{i,j,i',j'})$, is an $(N_s \times N_r) \times (N_s \times N_r)$ matrix;
One element of the diagonal terms of the Q matrix represents the cost of keypoint pair $(s_i, r_j)$.
One element of the off-diagonal terms of the Q matrix represents the cost of the combination of keypoint pair $(s_i, r_j)$ and keypoint pair $(s_{i'}, r_{j'})$.
The number of elements in the decision variable, $\mathbf{b}$, is $N_s \times N_r$.
Each element of the decision variable, $b_{i,j}$ $(i = 0 \ldots N_s - 1, j = 0 \ldots N_r - 1)$, is a binary variable that takes 1 if $s_i$ in the source image and $r_j$ in the reference image form a keypoint pair, and 0 otherwise, as shown in the following equation.
Note that $b_{i,j}^2 = b_{i,j}$.
\begin{align}
    b_{i,j} = 
    \begin{cases}
        1, & \text{if $s_i$ and $r_j$ are a keypoint pair} \\
        0, & \text{if $s_i$ and $r_j$ are not a keypoint pair}
    \end{cases}
\end{align}

Furthermore, when considering feature cost, geometry cost, co-occurrence cost, and constraints, the objective function $H_{\text{total}}(\mathbf{b})$ can be written as follows.
\begin{align}
    \begin{split}
    H_{\text{total}}(\mathbf{b})
         = {\alpha}_{f} \, H_{\text{feature}}(\mathbf{b})
         + {\alpha}_{g} \, H_{\text{geometry}}(\mathbf{b}) \\
         + {\alpha}_{c} \, H_{\text{cooccurrence}}(\mathbf{b}) 
         + {\alpha}_{p} \, H_{\text{penalty}}(\mathbf{b})
    \end{split}
    \label{eq:h-total}
\end{align}
$H_{\text{total}}$ is composed of a linear combination of:
the feature function, $H_{\text{feature}}$, corresponding to the feature cost of a single keypoint pair,
the geometry function, $H_{\text{geometry}}$, corresponding to the geometry cost of a single keypoint pair,
the co-occurrence function, $H_{\text{cooccurrence}}$, corresponding to the co-occurrence cost of two keypoint pairs, and
the penalty function, $H_{\text{penalty}}$, corresponding to the constraint condition elements of two keypoint pairs.
${\alpha}_{f}$, ${\alpha}_{g}$, ${\alpha}_{c}$, ${\alpha}_{p}$ are the weight coefficients of $H_{\text{feature}}$, $H_{\text{geometry}}$, $H_{\text{cooccurrence}}$, $H_{\text{penalty}}$, respectively.
$H_{\text{feature}}$ and $H_{\text{geometry}}$ are linear functions including the diagonal components of the Q matrix.
$H_{\text{cooccurrence}}$ and $H_{\text{penalty}}$ are quadratic functions including the off-diagonal components of the Q matrix.

$H_{\text{feature}}$ is expressed by
\begin{align}
    H_{\text{feature}}(\mathbf{b}) &= \sum_{\substack{0 \le i < N_s \\ 0 \le j < N_r}} F_{i,j} \, b_{i,j}.
\end{align}
$F_{i,j}$ is a real number representing the feature cost of keypoint pair $(s_i, r_j)$.
The smaller the distance between the features ${\mathbf{f}}^{S}_{i}$ and ${\mathbf{f}}^{R}_{j}$ of the two keypoints $s_i$ and $r_j$ of the keypoint pair, the smaller the feature cost $F_{i,j}$.
$F_{i,j}$ is defined as follows.
\begin{align}
    F_{i,j} &= \frac{dist_{\text{feature}} ({\mathbf{f}}^{S}_{i}, {\mathbf{f}}^{R}_{j})}{256} - {\beta}_{f}
\end{align}
$dist_{\text{feature}}({\mathbf{f}}^{S}_{i}, {\mathbf{f}}^{R}_{j})$ is a function that computes the feature distance between the feature descriptor ${\mathbf{f}}^{S}_{i}$ of keypoint $s_i$ and the feature descriptor ${\mathbf{f}}^{R}_{j}$ of keypoint $r_j$. The feature distance is computed as the Hamming distance between ${\mathbf{f}}^{S}_{i}$ and ${\mathbf{f}}^{R}_{j}$.
$F_{i,j}$ is normalized to $[- {\beta}_{f}, 1.0 - {\beta}_{f}]$ by dividing the feature distance by 256, which is the dimensionality of the ORB descriptor.
${\beta}_{f}$ is a bias term.

$H_{\text{geometry}}$ is expressed by
\begin{align}
    H_{\text{geometry}}(\mathbf{b}) &= \sum_{\substack{0 \le i < N_s \\ 0 \le j < N_r}} G_{i,j} b_{i,j}.
\end{align}
$G_{i,j}$ is a real number representing the geometry cost of keypoint pair $(s_i, r_j)$.
Let ${\hat{s}}_{i}$ denote the projection of source image keypoint $s_i$ onto the reference image. The geometric cost $G_{i,j}$ decreases as ${\hat{s}}_{i}$ approaches keypoint $r_j$.
The geometric distance is defined as the distance between the position ${\mathbf{\hat{p}}}^{S}_{i}$ of ${\hat{s}}_{i}$ on the reference image and the position ${\mathbf{p}}^{R}_{j}$ of keypoint $r_j$ on the reference image.
When the self-pose is unknown, the exact projected position cannot be determined, so the position ${\mathbf{\hat{p}}}^{S}_{i}$ of source image keypoint $s_i$ on the reference image is obtained by estimation.
The method for computing the estimated position ${\mathbf{\hat{p}}}^{S}_{i}$ is described in Section \ref{sec:eval-methodology}.
$G_{i,j}$ is defined as follows.
\begin{align}
    G_{i,j} &= \frac{dist_{\text{geometry}}({\mathbf{\hat{p}}}^{S}_{i}, {\mathbf{p}}^{R}_{j})}{64} - {\beta}_{g} \label{eq:geometry_distance}
\end{align}
$dist_{\text{geometry}}({\mathbf{\hat{p}}}^{S}_{i}, {\mathbf{p}}^{R}_{j})$ is a function that computes the geometric distance $|{\mathbf{\hat{p}}}^{S}_{i} - {\mathbf{p}}^{R}_{j}|$ between the estimated position ${\mathbf{\hat{p}}}^{S}_{i}$ and the position ${\mathbf{p}}^{R}_{j}$.
$G_{i,j}$ is normalized to $[- {\beta}_{g}, 1.0 - {\beta}_{g}]$ by dividing the geometric distance by 64, which is the distance threshold used in ORB-SLAM3's keypoint pair candidate search \cite{slam01-3_campos2021orb}.
${\beta}_{g}$ is a bias term.

$H_{\text{cooccurrence}}$ is expressed by
\begin{align}
    H_{\text{cooccurrence}}(\mathbf{b}) &= \sum_{\substack{0 \le i < N_s \\ 0 \le j < N_r}} \sum_{\substack{0 \le i' < N_s \\ 0 \le j' < N_r}} C_{i,j,i',j'} b_{i,j} b_{i',j'}.
\end{align}
When an ideal co-occurrence cost, $C_{i,j,i',j'}$, is defined, $H_{\text{cooccurrence}}$ takes its minimum value when all most likely combinations of two keypoint pairs are selected.
To consider the pairwise co-occurrence likelihood of two keypoint pairs, this paper assumes the following constraints on the source and reference images: (a) most objects are stationary objects, and the targets of keypoint matching are also stationary objects; (b) camera translation and rotation between the source image and the reference image are small.
Based on these assumptions, it can be assumed that keypoints that are close to each other in the source image are also close to each other in the reference image.
The co-occurrence cost reflecting the pairwise co-occurrence likelihood of two keypoint pairs has the following elements.
\begin{enumerate}
    \item Distance between keypoints $s_i$ and $s_{i'}$: We assume that the smaller the geometric distance between keypoints $s_i$ and $s_{i'}$ in the source image of the two keypoint pairs, the greater the possibility that they are also close in the reference image. \label{enum:coocc:01}
    \item Direction of ${\mathbf{v}}_{i,j}$ and ${\mathbf{v}}_{i',j'}$ (cosine similarity): We assume that the more similar the directions of the two keypoint pairs, the greater the possibility that they simultaneously hold. \label{enum:coocc:02}
    \item Magnitude of ${\mathbf{v}}_{i,j}$ and ${\mathbf{v}}_{i',j'}$ : We assume that the closer the magnitudes of the vectors of the two keypoint pairs, the greater the possibility that they simultaneously hold. This is more effective when the directions of the two keypoint pairs are similar. \label{enum:coocc:03}
\end{enumerate}
The proposed method defines $C_{i,j,i',j'}$ by combining these elements as follows.
\begin{equation}
    C_{i,j,i',j'} = 
    \begin{cases}
        f({\mathbf{v}}_{i,j}, {\mathbf{v}}_{i',j'}), & \text{if $dist_{\text{geometry}}({\mathbf{p}}^{S}_{i}$, ${\mathbf{p}}^{S}_{i'}) < threshold$} \\
        0, & \text{otherwise}
    \end{cases}
\end{equation}
\begin{align}
    f({\mathbf{x}}, {\mathbf{y}}) &= - {\text{cossim}}({\mathbf{x}}, {\mathbf{y}}) \, {\text{ratio}}({\mathbf{x}}, {\mathbf{y}}) \\
    {\text{cossim}}({\mathbf{x}}, {\mathbf{y}}) &= \frac{{\mathbf{x}} \cdot {\mathbf{y}}}{|{\mathbf{x}}| |{\mathbf{y}}|}  \\
    {\text{ratio}}({\mathbf{x}}, {\mathbf{y}}) &= \frac{\min(|{\mathbf{x}}|, |{\mathbf{y}}|)}{\max(|{\mathbf{x}}|, |{\mathbf{y}}|)}
\end{align}
When the distance between two keypoints $s_i$ and $s_{i'}$ in the source image is less than $threshold$,
$C_{i,j,i',j'}$ uses the cosine similarity, ${\text{cossim}}({\mathbf{v}}_{i,j}, {\mathbf{v}}_{i',j'})$, of the two keypoint pairs multiplied by
the ratio, ${\text{ratio}}({\mathbf{v}}_{i,j}, {\mathbf{v}}_{i',j'})$, of the magnitudes of the two keypoint pair vectors, ${\mathbf{v}}_{i,j}$ and ${\mathbf{v}}_{i',j'}$.
When the distance between the two keypoints exceeds $threshold$, $C_{i,j,i',j'}$ is set to 0.
The value of this $threshold$ is set to 64, based on the distance threshold used in keypoint pair candidate search \cite{slam01-3_campos2021orb}.

In local feature matching, there is a constraint that a keypoint in the source image can form a pair with at most one keypoint in the reference image, and a keypoint in the reference image can form a pair with at most one keypoint in the source image.
$H_{\text{penalty}}$ is the penalty function representing this constraint.
\begin{align}
    H_{\text{penalty}}(\bf{b}) &= \sum_{\substack{0 \le i < N_s \\ 0 \le j < N_r}} \sum_{\substack{0 \le i' < N_s \\ 0 \le j' < N_r}} P_{i,j,i',j'} b_{i,j} b_{i',j'}
\end{align}
$P_{i,j,i',j'}$ is defined as follows.
\begin{equation}
    P_{i,j,i',j'} = 
    \begin{cases}
        1 & (i = i' \land j \neq j') \lor (i \neq i' \land j = j') \\
        0 & otherwise
    \end{cases}
\end{equation}

\subsection{Reducing Number of Required Decision Variables} \label{sec:lfmising-spinreduction}

The size of an Ising machine, characterized by the number of available spins and couplings, determines the maximum number of decision variables (bits) that can be accommodated in an optimization problem.
Particularly when assuming deployment in embedded applications, available hardware resources are limited also by constraints such as circuit scale and power consumption.
Therefore, in embedded applications, reducing the number of bits required to solve optimization problems is especially important.

As shown in Section \ref{sec:lfmising-formulation}, the number of bits required by the proposed method is $(N_s \times N_r)$.
When 500 keypoints are detected per image, the required number of bits becomes 250,000.
Since digital circuit-based Ising machines used in embedded applications \cite{tatsumura2025enhancing} and \cite{motcite60_hidaka2023correlation} support 512 to 2048 spins, this number of bits (250,000) would be too large for use in embedded devices.
To efficiently execute the proposed method with hardware resources available for embedded applications, the required number of bits needs to be reduced to a few thousand.

As a method for reducing the required number of bits while maintaining the number of keypoints to be searched, we propose a method that selectively chooses candidate keypoints in the reference image for each source image keypoint.
Since keypoint pairs with large feature distances are less likely to represent correct correspondences,
excluding these keypoint pairs from the candidates leaves only candidates with a higher probability of correct correspondences and thereby reduces the number of bits.
The procedure for selecting $K$ reference image keypoint candidates for each source image keypoint $s_i (i = 0, \ldots, N_s - 1)$ is as follows.
\begin{enumerate}
    \item Compute the estimated position ${\mathbf{\hat{p}}}^{S}_{i}$ of source image keypoint $s_i$ on the reference image. \label{enum:spinreduce_step1}
    \item Select reference image keypoints $r_j (j = 0, \ldots, N_r - 1)$ for which $dist_{\text{geometry}}({\mathbf{\hat{p}}}^{S}_{i}, {\mathbf{p}}^{R}_{j})$ is below the threshold. \label{enum:spinreduce_step2}
    \item From the reference image keypoints $r_j$ selected in step \ref{enum:spinreduce_step2}, select the top $K$ keypoints with the smallest $dist_{\text{feature}}({\mathbf{f}}^{S}_{i}, {\mathbf{f}}^{R}_{j})$ (or smallest $F_{i,j}$) as the final candidates. \label{enum:spinreduce_step3}
\end{enumerate}

Using this selection method, the required number of bits can be reduced to $(N_s \times K)$.
When 500 keypoints are detected per image, setting $K = 4$ allows the required number of bits to be dramatically reduced from 250,000 to 2,000.
By selecting candidate keypoint pairs, effective local feature matching can be realized even with limited hardware resources.

\subsection{Local Feature Matching Based on Co-occurrence-Aware Assignment} \label{sec:lfmising-sub}
\input{src/fig_yamad5.tex}

Figure \ref{fig:yamad5_qubolfm} shows the local feature matching pipeline using Co-occurrence-aware Keypoint Matching, which consists of keypoint detection, feature extraction, and keypoint matching, as in the conventional method.
In the proposed method, only the conventional keypoint matching stage is replaced with Co-occurrence-aware Keypoint Matching.
The keypoint matching generates the Q matrix using the position information and feature descriptors of the keypoints, solves the QAP represented by the objective function $H_{\text{total}}$, and outputs the solution as the correspondence of keypoint pairs.
In this way, local feature matching utilizing the pairwise co-occurrence likelihood of two keypoint pairs can be realized.

\subsection{Visual SLAM System with Local Feature Matching using Co-occurrence-Aware Quadratic Assignment} \label{sec:slamising}

We describe a system that uses Co-occurrence-aware Keypoint Matching in feature-based Visual SLAM, which is shown in Figure \ref{fig:yamad1_concept}.
The proposed system consists of three processes: Tracking, Local Mapping, and Loop Closing, similar to the conventional Visual SLAM described in Section \ref{sec:slam}.
In the Tracking process of the proposed system, the local feature matching is replaced with one using Co-occurrence-aware Keypoint Matching presented in Section \ref{sec:lfmising-sub}.
The other processes are the same as those of conventional Visual SLAM.

\section{Evaluation} \label{sec:evaluation}

\subsection{Methodology} \label{sec:eval-methodology}

This section evaluates the effectiveness of Co-occurrence-aware Keypoint Matching presented in Section \ref{sec:lfmising}.
The objective is to demonstrate that pose estimation accuracy for scenes where objects of the same shape are continuously arranged without remarkable landmarks is improved by using Co-occurrence-aware Keypoint Matching for the keypoint matching in feature-based Visual SLAM.
The publicly available KITTI dataset \cite{Geiger2012CVPR} is used for accuracy evaluation of Visual SLAM self-pose estimation.
Since scenes where objects of the same shape are continuously arranged without remarkable landmarks are included in Sequence 01 of the KITTI dataset, Sequence 01 is used for evaluation.

As a supplementary evaluation of Visual SLAM self-pose estimation accuracy, we also evaluate whether using Co-occurrence-aware Keypoint Matching for a standalone local feature matcher improves correspondence point accuracy.
The HPatches dataset \cite{hpatches_2017_cvpr}, a publicly available dataset, is used for accuracy evaluation of local feature matching.
For the evaluation of local feature matching, the evaluation setup is designed to closely resemble that of the Visual SLAM evaluation environment.
Because some components cannot be implemented due to differences between the Visual SLAM environment and the standalone local feature matching environment, simplified implementations are used for certain functions, including the computation of ${\mathbf{\hat{p}}}^{S}_{i}$.

\paragraph{Accuracy Evaluation of Visual SLAM}

The feature-based Visual SLAM, a monocular camera configuration of ORB-SLAM3 \cite{slam01-3_campos2021orb}, is used as the baseline.
The proposed method replaces only the keypoint matching part of ORB-SLAM3 with the method presented in Section \ref{sec:lfmising-formulation}.
All other parts, including parameter settings, use those of ORB-SLAM3 as-is.
ORB-SLAM3 computes the estimated position ${\mathbf{\hat{p}}}^{S}_{i}$ on the reference image for each source image keypoint $s_i$ using past self-pose estimation information, which is used for selecting reference image keypoint candidates.
This value is also used for the estimated position ${\mathbf{\hat{p}}}^{S}_{i}$ used in computing the geometry cost $G_{i,j}$ of the proposed method.
The number of candidates, $K$, for the reduction method presented in Section \ref{sec:lfmising-spinreduction} is set to $\min({\lfloor}2048 / N_s{\rfloor}, 8)$.
The coefficients shown in Equation (\ref{eq:h-total}) are set as described in Appendix \ref{sec:append-param}:
${\alpha}_f = 2.0$, ${\alpha}_g = 8.0$, ${\alpha}_c = 8.0$, ${\alpha}_p = 64.0$, ${\beta}_f = 1.0$, ${\beta}_g = 0.03125$.

\paragraph{Accuracy Evaluation of Local Feature Matching} \label{sec:eval-method-lfm}

For the evaluation of local feature matching using the HPatches dataset \cite{hpatches_2017_cvpr}, an environment similar to the local feature matching used in ORB-SLAM3 is implemented using OpenCV \cite{opencv_library}.
The baseline uses \texttt{cv2.ORB} for keypoint detection and feature extraction, and \texttt{cv2.BFMatcher(crossCheck=True)} for NN search-based keypoint matching.
Since we confirmed that the maximum value of $N_s$ in ORB-SLAM3 is at most approximately 500 points except for initialization phases, the maximum number of keypoints for ORB detection is set to 500 points.
For the proposed method, the keypoint matching part is replaced from \verb/cv2.BFMatcher/ to Co-occurrence-aware Keypoint Matching presented in Section \ref{sec:lfmising-formulation}.
The number of candidates for the reduction method presented in Section \ref{sec:lfmising-spinreduction} is set to $K = 4$ according to the number of detected keypoints.
The coefficients shown in Equation (\ref{eq:h-total}) are set as described in Appendix \ref{sec:append-param}:
${\alpha}_f = 128.0$, ${\alpha}_g = 32.0$, ${\alpha}_c = 32.0$, ${\alpha}_p = 128.0$, ${\beta}_f = 0.25$, ${\beta}_g = 1.0$.

Since the evaluation of local feature matching uses only two images, the estimated position ${\mathbf{\hat{p}}}^{S}_{i}$ cannot be computed using self-pose estimation as in ORB-SLAM3.
Therefore, the following simplified method is used for estimating the position ${\mathbf{\hat{p}}}^{S}_{i}$.
\begin{enumerate}
\item Match source image keypoints $s_i (i=0 \ldots N_s - 1)$ and reference image keypoints $r_j (j=0 \ldots N_r - 1)$ using NN search (Baseline) to obtain keypoint pairs $(s_0, r_{s_0}), (s_1, r_{s_1}), \cdots$. \label{enum:sec:eval-method-step1}
\item Obtain four neighboring keypoints $s_{i0}, s_{i1}, s_{i2}, s_{i3}$ for the target source image keypoint $s_i$.
\item Obtain the positions ${\mathbf{p}}^{R}_{j0}, {\mathbf{p}}^{R}_{j1}, {\mathbf{p}}^{R}_{j2}, {\mathbf{p}}^{R}_{j3}$ of the keypoint pairs $r_{s0}, r_{s1}, r_{s2}, r_{s3}$ searched in step \ref{enum:sec:eval-method-step1} for the four neighboring keypoints $s_{i0}, s_{i1}, s_{i2}, s_{i3}$.
\item Compute the estimated value ${\mathbf{\hat{p}}}^{S}_{i}$ by linear interpolation from the coordinates of ${\mathbf{p}}^{R}_{j0}, {\mathbf{p}}^{R}_{j1}, {\mathbf{p}}^{R}_{j2}, {\mathbf{p}}^{R}_{j3}$.
\end{enumerate}

This method obtains estimated positions in a simplified manner by using the results of baseline keypoint matching.
Therefore, the accuracy of the proposed method may be influenced by the baseline (NN search) accuracy.

The evaluation compares a baseline method with the proposed method using four variants of keypoint matching methods based on the presence or absence of the geometry function and co-occurrence function.
The evaluated algorithms are listed below. Ising(F, G, C) corresponds to the same processing as the Visual SLAM evaluation target.
\begin{itemize}
    \item Baseline: NN search by feature distance
    \item Ising(F): Search with an Ising machine using an objective function including the feature function $H_{\text{feature}}({\mathbf{b}})$ and penalty function $H_{\text{penalty}}({\mathbf{b}})$
    \item Ising(F, C): Search with an Ising machine using an objective function including the feature function $H_{\text{feature}}({\mathbf{b}})$, co-occurrence function $H_{\text{cooccurrence}}({\mathbf{b}})$, and penalty function $H_{\text{penalty}}({\mathbf{b}})$
    \item Ising(F, G): Search with an Ising machine using an objective function including the feature function $H_{\text{feature}}({\mathbf{b}})$, geometry function $H_{\text{geometry}}({\mathbf{b}})$, and penalty function $H_{\text{penalty}}({\mathbf{b}})$
    \item Ising(F, G, C): Search with an Ising machine using an objective function including the feature function $H_{\text{feature}}({\mathbf{b}})$, geometry function $H_{\text{geometry}}({\mathbf{b}})$, co-occurrence function $H_{\text{cooccurrence}}({\mathbf{b}})$, and penalty function $H_{\text{penalty}}({\mathbf{b}})$
\end{itemize}

\subsection{Evaluation System} \label{sec:systemarch}

\input{src/fig_yamad6.tex}

Figure \ref{fig:yamad6_sysarch} shows the evaluation system used to conduct the comparative experiments described in Section \ref{sec:eval-methodology}.
In this paper, we adopted an implementation using an Ising machine on a server with the primary objective of demonstrating the effectiveness of the proposed method, while also envisioning deployment in embedded applications for the future.

This evaluation system comprises a client that performs software processing and a server that performs optimization processing using an Ising machine.
The Ising machine on the server uses a Simulated Bifurcation Machine (SBM) \cite{motcite19_tatsumura2019fpga, motcite60_hidaka2023correlation} implemented on an FPGA. See Appendix \ref{sec:append-sbm} for details on Simulated Bifurcation.
The maximum number of spins that this Ising machine can handle is 2048.
The client and server are connected via 1Gb Ethernet.

Figure \ref{fig:yamad6_sysarch} also indicates the processing flow of local feature matching, which is as follows:
\begin{enumerate}
\item The client performs keypoint detection and feature extraction for the source and reference images, and generates the Q matrix for keypoint matching.
\item The client sends the Q matrix to the server.
\item The server solves the QAP represented by the Q matrix received from the client and sends the assignment result to the client.
\item The client receives the assignment result from the server and determines the keypoint pairs.
\end{enumerate}
The server handles only the combinatorial optimization processing by the Ising machine, and all other processing is performed in software on the client.
Figure \ref{fig:yamad6_sysarch} shows the system architecture for the evaluation of local feature matching using the HPatches dataset \cite{hpatches_2017_cvpr}.
The system architecture for Visual SLAM can be implemented by replacing keypoint matching in the Visual SLAM software on the client with Co-occurrence-aware Keypoint Matching and by processing the QAP of keypoint matching on the server.

\subsection{Evaluation for Local Feature Matching} \label{sec:evaluation-hpatches}

This section describes the accuracy evaluation of local feature matching using Co-occurrence-aware Keypoint Matching as presented in Section \ref{sec:eval-methodology}.

\paragraph{Dataset}
The publicly available HPatches dataset \cite{hpatches_2017_cvpr} was used as the evaluation data.
The HPatches dataset contains sequences consisting of 5 pairs of images each from 57 illumination change datasets and 59 viewpoint change datasets.
In this study, the viewpoint change datasets were used for evaluation.

\paragraph{Metrics}
The evaluation metrics used were: the total number of matched keypoint pairs ($N_{\text{matches}}$), the number of correctly matched keypoint pairs (TP, true positives), the number of incorrectly matched keypoint pairs (FP, false positives), and the accuracy of correct correspondences (Precision).
A correctly matched keypoint pair was determined as one where the error between the matched keypoint and its ground truth was within 5 pixels.

\paragraph{Result}
\input{src/tab_yamad2.tex}
\input{src/fig_yamad7.tex}

Table \ref{tab:yamad2_hpatches} lists the evaluation results in terms of $N_{\text{matches}}$, TP, FP, and Precision. Each evaluation metric shows the average value across all datasets.
In the figures and tables in this paper, $(\uparrow)$ means higher is better, and $(\downarrow)$ means lower is better.

When only the feature function was used with the Ising machine (Ising(F)), the accuracy was equivalent to the baseline.
This is a reasonable result since the information used for correspondence search is identical.
It was confirmed that using the co-occurrence function (Ising(F, C)) improved TP and Precision, and suppressed FP.
Improvement in Precision was also confirmed when incorporating geometry cost (Ising(F, G)).
Although the number of TP decreased, the reduction in FP was even greater, leading to improvement in Precision.
The method using both the geometry function and co-occurrence function (Ising(F, G, C)) confirmed an improvement of approximately 8 percentage points  (\textbf{pp}) in Precision and approximately 9\% in TP count compared to the baseline.

Figure \ref{fig:yamad7_hpatches} shows sample local feature matching results.
Correct and incorrect correspondences are shown with green and red lines, respectively.
It was confirmed that the proportion of correctly matched points increased by considering the geometry function and co-occurrence function.
Because the geometry function uses baseline keypoint matching results to estimate ${\mathbf{\hat{p}}}^{S}_{i}$ (see Section \ref{sec:eval-methodology}-b),
for cases like v\_artisans\_6 where the baseline accuracy is low, using the geometry function increased the proportion of incorrectly matched keypoint pairs.
On the other hand, when using the co-occurrence function, it was confirmed that incorrectly matched keypoint pairs were generally suppressed, although the degree of improvement varied.
In particular, it was confirmed that using the co-occurrence function suppressed incorrectly matched keypoint pairs in images containing many objects with similar shapes, such as v\_sample\_4 and v\_color\_2.

\subsection{Evaluation for Visual SLAM} \label{sec:evaluation-slam}

This section describes the accuracy evaluation of Visual SLAM using Co-occurrence-aware Keypoint Matching as presented in Section \ref{sec:slamising}.

\paragraph{Dataset}
The publicly available KITTI dataset \cite{Geiger2012CVPR} was used as the evaluation data for SLAM.
To confirm the accuracy improvement effect of local feature matching, sequences without loop closure that do not execute global map optimization (Global Bundle Adjustment) were used.

\paragraph{Metrics}
The evaluation metrics used were the Root Mean Square Error (RMSE) of Absolute Pose Error (APE) and Relative Pose Error (RPE) evaluated by evo \cite{grupp2017evo} in addition to the average total number of matched keypoint pairs per frame ($N_{\text{matches}}$).
The baseline ORB-SLAM3 and the system using Co-occurrence-aware Keypoint Matching were executed 5 times, and the median of the results of the 5 runs was used as the evaluation value \cite{slam01_mur2015orb, slam01-2_mur2017orb}.

\paragraph{Result}
\input{src/tab_yamad3.tex}
\input{src/fig_yamad8.tex}
\input{src/fig_yamad9.tex}

Table \ref{tab:yamad3_kitti} lists RMSE of APE and RPE as well as $N_{\text{matches}}$.
Additionally, Repetitive indicates whether the scene contains objects of the same shape repeatedly arranged with few other keypoints available as landmarks.
In sequences without Repetitive scenes, the baseline and the proposed method show equivalent performance.
In these sequences, the baseline already achieves good accuracy for both APE and RPE, leaving little room for improvement by the proposed method.
In contrast, in Sequence01, which includes Repetitive scenes, the baseline shows worse APE and RPE values than in other sequences.
In this sequence, the proposed method confirmed improvements in both APE and RPE compared to the baseline.

Figure \ref{fig:yamad8_kittigraph} shows the improvement rate of APE and RPE of the proposed method relative to the baseline.
In sequences without Repetitive scenes (Sequences 03, 04, 08, and 10), the baseline already attains good accuracy, and the improvement factors are around 1.00-fold for both APE and RPE.
By contrast, in Sequence01, which includes Repetitive scenes, the improvement factors show marked gains: 3.78-fold for APE and 2.85-fold for RPE.

Figure \ref{fig:yamad9_kitti} shows the local feature matching results in scenes of the sequences used for evaluation.
The left side shows the local feature matching results of the baseline (ORB-SLAM3), and the right side shows the results of the proposed method.
The figures show input frames (source images), and the arrows indicate the motions from the keypoints in the previous frames (reference images) to their corresponding keypoints in the input frames.
The arrow colors indicate the motion direction and correspond to the color chart shown at the upper right.
For correctly matched keypoint pairs, keypoint pair motion is expected to be consistent with optical flow, which represents object motion between frames.
When objects are stationary, this motion is determined by camera motion.
If the camera moves forward, the motion is expected to be radially outward from the vanishing point.
If the camera rotates, the motion is expected to follow the rotation direction.
Because stationary objects in a frame move coherently with camera motion, the motions of nearby stationary objects are expected to be similar.

In the baseline, as shown on the left side of Figure \ref{fig:yamad9_kitti}, keypoint pairs with motion different from nearby keypoint pairs are observed in all scenes, regardless of whether Repetitive scenes are present.
Although many keypoint pairs in the images follow optical flow consistent motion, some keypoint pairs show motion inconsistent with optical flow, and these can be regarded as incorrect matches.
Such incorrect keypoint pairs are observed on same shape objects such as guardrail poles, windows, and trees.
In non-Repetitive scenes, the motion of incorrect keypoint pairs varies with respect to optical flow, and no regular pattern is observed.
In Sequence01, which is a Repetitive scene, keypoint pair motion opposite to optical flow is observed.
This occurs because same shape objects such as poles and lane markings are aligned linearly, and when incorrect keypoints are matched, the resulting keypoint pair motion becomes reversed.
If keypoint pair motion opposite to optical flow is included in self-pose estimation, it may be misinterpreted as motion in the opposite direction.
In particular, Sequence01 has few other landmark keypoints, so opposite direction keypoint pairs can have a larger adverse impact on self-pose estimation.

For the proposed method, as shown on the right side of Figure \ref{fig:yamad9_kitti}, these incorrect keypoint pairs are suppressed.
In the baseline, keypoint pairs with motion different from nearby keypoint pairs produce high co-occurrence costs.
Especially in Sequence01, a Repetitive scene, incorrect keypoint-pair motion is opposite to optical flow and thus incurs particularly high co-occurrence costs.
Therefore, in the proposed method, incorrectly matched keypoint pairs with motion different from nearby keypoint pairs are considered to be suppressed because they receive high co-occurrence costs.
Suppressing incorrect keypoint pairs is expected to improve self-pose estimation accuracy.

Supplementary material provides videos of local feature matching results for all scenes in the evaluation sequences.
In the videos, the upper part shows the baseline (ORB-SLAM3), and the lower part shows the proposed method.
As in Figure \ref{fig:yamad9_kitti}, arrows indicate directions from keypoints in past frames to corresponding keypoints in input frames.
In the supplementary videos, the occurrence of incorrect keypoint pairs in the baseline can be observed more clearly.
These incorrect keypoint pairs are not temporally consistent across frames and can be seen to match different keypoints from frame to frame.
This indicates that frame-wise self-pose estimation results may become unstable.
By contrast, the videos also confirm that the proposed method suppresses the occurrence of incorrect keypoint pairs.
By suppressing such incorrect keypoint pairs, the proposed method is expected to improve the stability of self-pose estimation.

\section{Conclusion} \label{sec:conclusion}

We proposed Co-occurrence-aware Keypoint Matching considering the co-occurrence likelihood of two keypoint pairs with the objective of improving pose estimation accuracy in SLAM, especially in challenging scenes where multiple objects of the same shape are continuously arranged without remarkable background landmarks.
We formulated the proposed Co-occurrence-aware Keypoint Matching as QUBO for solving on an Ising machine and used a digital circuit-based Ising machine while taking into account deployment in embedded applications. 
In quantitative evaluation of local feature matching on the HPatches dataset, the proposed method improved matching accuracy by 8\% compared to conventional NN search.
In the quantitative evaluation of pose estimation in SLAM on the KITTI dataset, the Visual SLAM system in which the keypoint matching of ORB-SLAM3 was replaced with the proposed method achieved APE improvement by 3.78-fold and RPE improvement by 2.85-fold in a challenging sequence (Sequence 01) where multiple objects of the same shape are continuously arranged without remarkable landmarks. In the other evaluated sequences, the proposed method showed broadly comparable performance, although some sequences exhibited moderate degradation in APE or RPE.

Future work includes algorithmic improvements to the proposed method aimed at further enhancing pose estimation accuracy in Visual SLAM.
For example, deep learning-based methods for local feature matching have been proposed, and further accuracy improvements can be expected by combining these with the proposed method.
Since the proposed method suppresses incorrect matches, simplification of processing introduced to enhance robustness of Visual SLAM may be possible, which can lead to further speedup.
For example, computational cost reduction can be expected by simplifying outlier removal processing of keypoint pairs in pose estimation.

Additionally, implementation on embedded systems is also envisioned, targeting deployment on autonomous driving vehicles, which is a major application of Visual SLAM.
In this paper, we focused on accuracy evaluation in local feature matching and Visual SLAM, and implemented a server-client model that separates the software processing part and the Ising machine processing part that solves the QUBO.
Communication latency between the server and client in this system becomes a challenge because embedded systems require real-time performance.
To implement an embedded system, it is necessary to construct a system where the software processing part and the Ising machine processing part are tightly coupled, as shown in our related work \cite{tatsumura2025enhancing}.

%% file: src/fig_yamad1.tex
\ifdefined\draft
\begin{figure*}[!ht]
\centering
\includegraphics[scale=0.20]{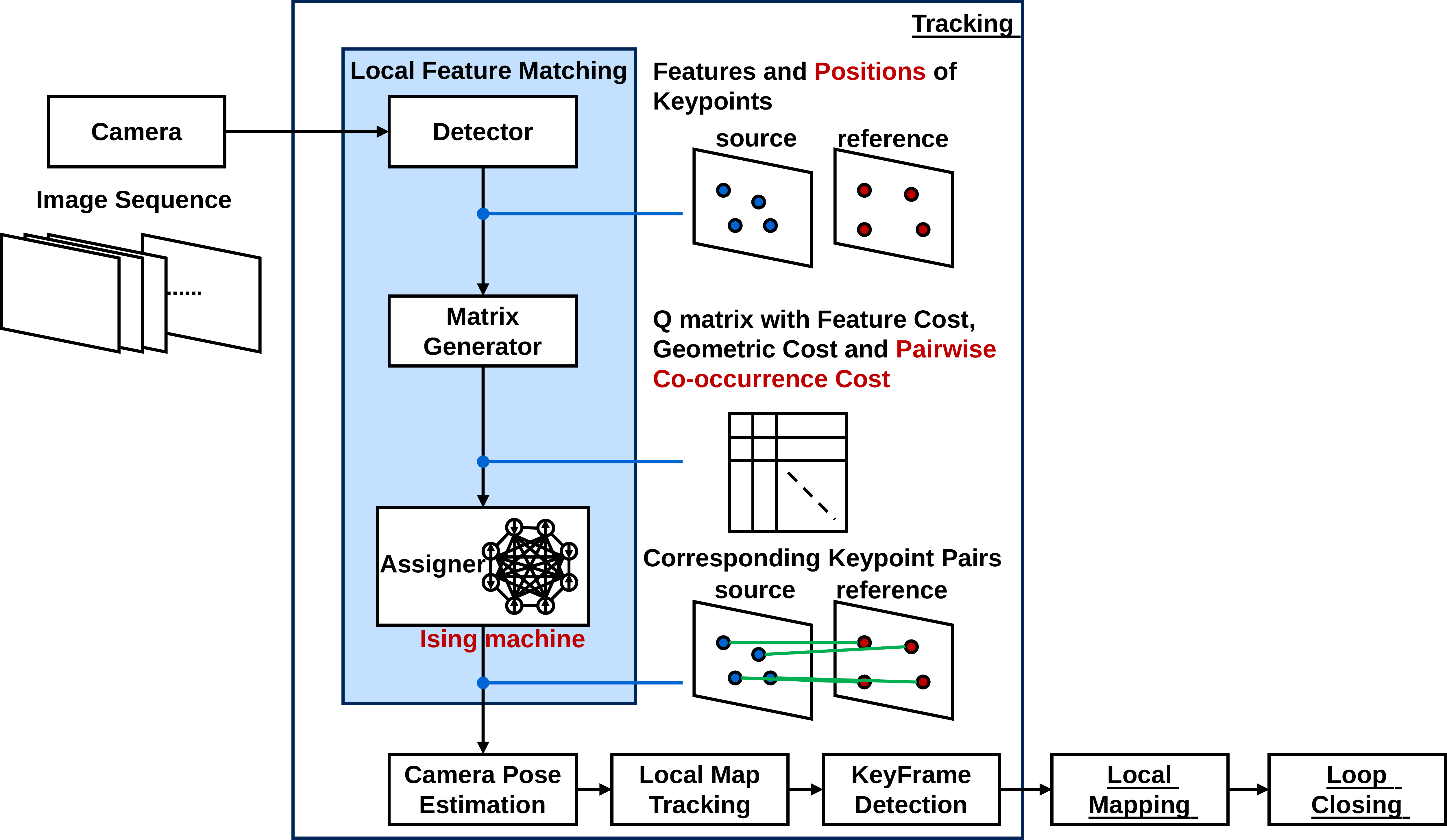}
\caption{Visual Simultaneous Localization and Mapping (Visual SLAM) system using an Ising machine for local feature matching.
The detector identifies keypoints in the camera images and extracts their feature descriptors. 
Throughout this paper, the blue dots and red dots indicate keypoints in the source image and the reference image, respectively.
The Matrix Generator calculates the cost matrix, which consists of feature cost, geometric cost, and pairwise co-occurrence cost, from the features and positions of the keypoints.
The assigner using the Ising machine solves the combinatorial optimization problem represented by the cost matrix and determines the keypoint pairs.
The subsequent processing (camera-pose estimation, local map tracking, keyframe detection, local mapping, and loop closing) can be performed in the same manner as in baseline Visual SLAM systems such as ORB-SLAM \cite{slam01_mur2015orb, slam01-2_mur2017orb, slam01-3_campos2021orb}.
}
\label{fig:yamad1_concept}
\end{figure*}
\else
\Figure[t!](topskip=0pt, botskip=0pt, midskip=0pt)[scale=0.20]{figure/yamad1.pdf}
{ \textbf{Visual Simultaneous Localization and Mapping (Visual SLAM) system using an Ising machine for local feature matching.
The detector identifies keypoints in the camera images and extracts their feature descriptors. 
Throughout this paper, the blue dots and red dots indicate keypoints in the source image and the reference image, respectively.
The Matrix Generator calculates the cost matrix, which consists of feature cost, geometric cost, and pairwise co-occurrence cost, from the features and positions of the keypoints.
The assigner using the Ising machine solves the combinatorial optimization problem represented by the cost matrix and determines the keypoint pairs.
The subsequent processing (camera-pose estimation, local map tracking, keyframe detection, local mapping, and loop closing) can be performed in the same manner as in baseline Visual SLAM systems such as ORB-SLAM \cite{slam01_mur2015orb, slam01-2_mur2017orb, slam01-3_campos2021orb}.
}
\label{fig:yamad1_concept}}
\fi

%% file: src/fig_yamad2.tex
\ifdefined\draft
\begin{figure*}[!ht]
\centering
\includegraphics[scale=0.20]{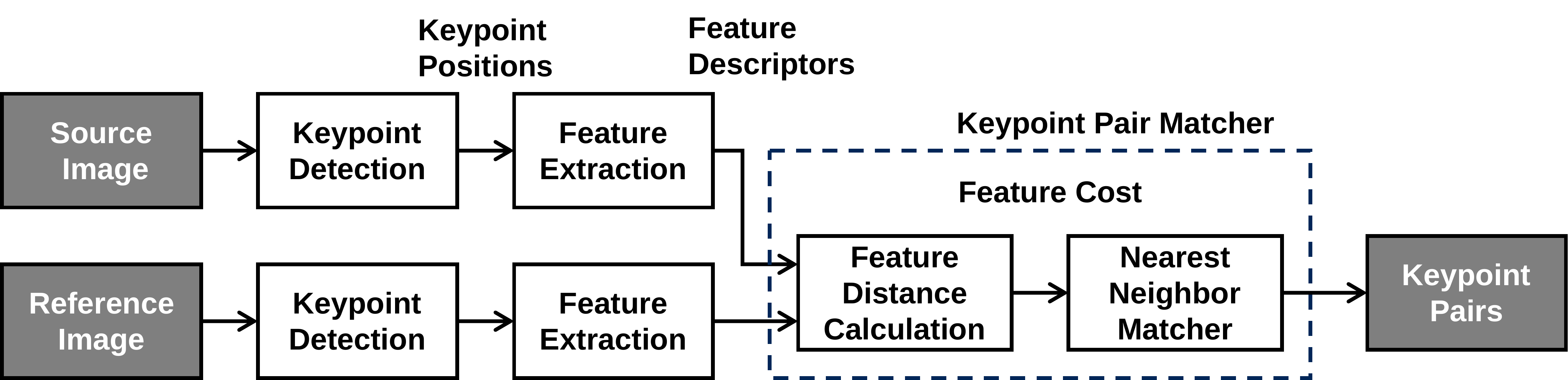}
\caption{Block diagram of conventional local feature matching.}
\label{fig:yamad2_lfm}
\end{figure*}
\else
\Figure[t!](topskip=0pt, botskip=0pt, midskip=0pt)[scale=0.20]{figure/yamad2.pdf}
{ \textbf{Block diagram of conventional local feature matching.}
\label{fig:yamad2_lfm}}
\fi

%% file: src/fig_yamad3.tex
\ifdefined\draft
\begin{figure*}[!ht]
\centering
\includegraphics[scale=0.20]{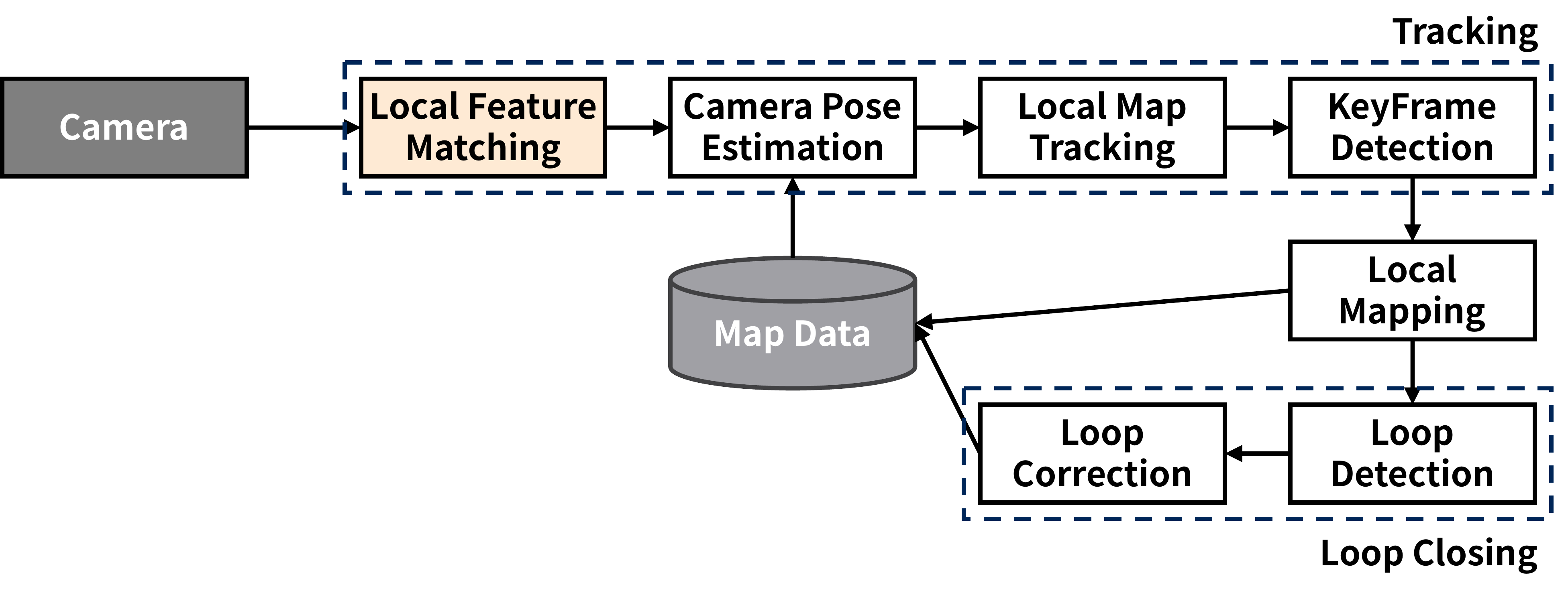}
\caption{Block diagram of feature-based monocular-camera Simultaneous Localization and Mapping modified from ORB-SLAM \cite{slam01_mur2015orb}.}
\label{fig:yamad3_slam}
\end{figure*}
\else
\Figure[t!](topskip=0pt, botskip=0pt, midskip=0pt)[scale=0.20]{figure/yamad3.pdf}
{ \textbf{Block diagram of feature-based monocular-camera Simultaneous Localization and Mapping modified from ORB-SLAM \cite{slam01_mur2015orb}.}
\label{fig:yamad3_slam}}
\fi

%% file: src/tab_yamad1.tex
\begin{table*}[tb]
\centering
\caption{\textbf{List of symbols and definitions.}} \label{table:03}
\setlength{\tabcolsep}{3pt}
\begin{tabular}{ll}
    \toprule
    \textbf{Symbol} & \textbf{Definition} \\
    \midrule
    $N_s$       & Number of keypoints in the source image    \\
    $N_r$       & Number of keypoints in the reference image \\
    $i$, $i'$   & Index of a keypoint in the source image    \\
    $j$, $j'$   & Index of a keypoint in the reference image \\
    $s_{i}$            & Keypoint indexed $i$ in the source image           \\
    ${\hat{s}}_{i}$    & Estimated position of $s_{i}$ in the reference image           \\
    $r_{j}$            & Keypoint indexed $j$ in the reference image        \\
    ${\mathbf{f}}^{S}_{i}$ & Feature descriptor of keypoint $s_i$ in the source image \\
    ${\mathbf{f}}^{R}_{j}$ & Feature descriptor of keypoint $r_j$ in the reference image \\
    ${\mathbf{p}}^{S}_{i}$ & Pixel position of keypoint $s_i$ in the source image    \\
    ${\mathbf{\hat{p}}}^{S}_{i}$ & Estimated pixel position of keypoint $s_i$ in the reference image    \\
    ${\mathbf{p}}^{R}_{j}$ & Pixel position of keypoint $r_j$ in the reference image \\
    ${\mathbf{v}}_{i,j}$ & Motion vector from ${\mathbf{p}}^{S}_{i}$ to ${\mathbf{p}}^{R}_{j}$ \\
    $(s_i, r_j)$ & Pair consisting of keypoints $s_i$ and $r_j$ \\
    $((s_i, r_j), (s_{i'}, r_{j'}))$ & Pair of keypoint pairs $(s_i, r_j)$ and $(s_{i'}, r_{j'})$ \\
    \bottomrule
\end{tabular}
\label{tab:yamad1_def}
\end{table*}

%% file: src/fig_yamad4.tex
\ifdefined\draft
\begin{figure*}[!ht]
\centering
\includegraphics[scale=0.20]{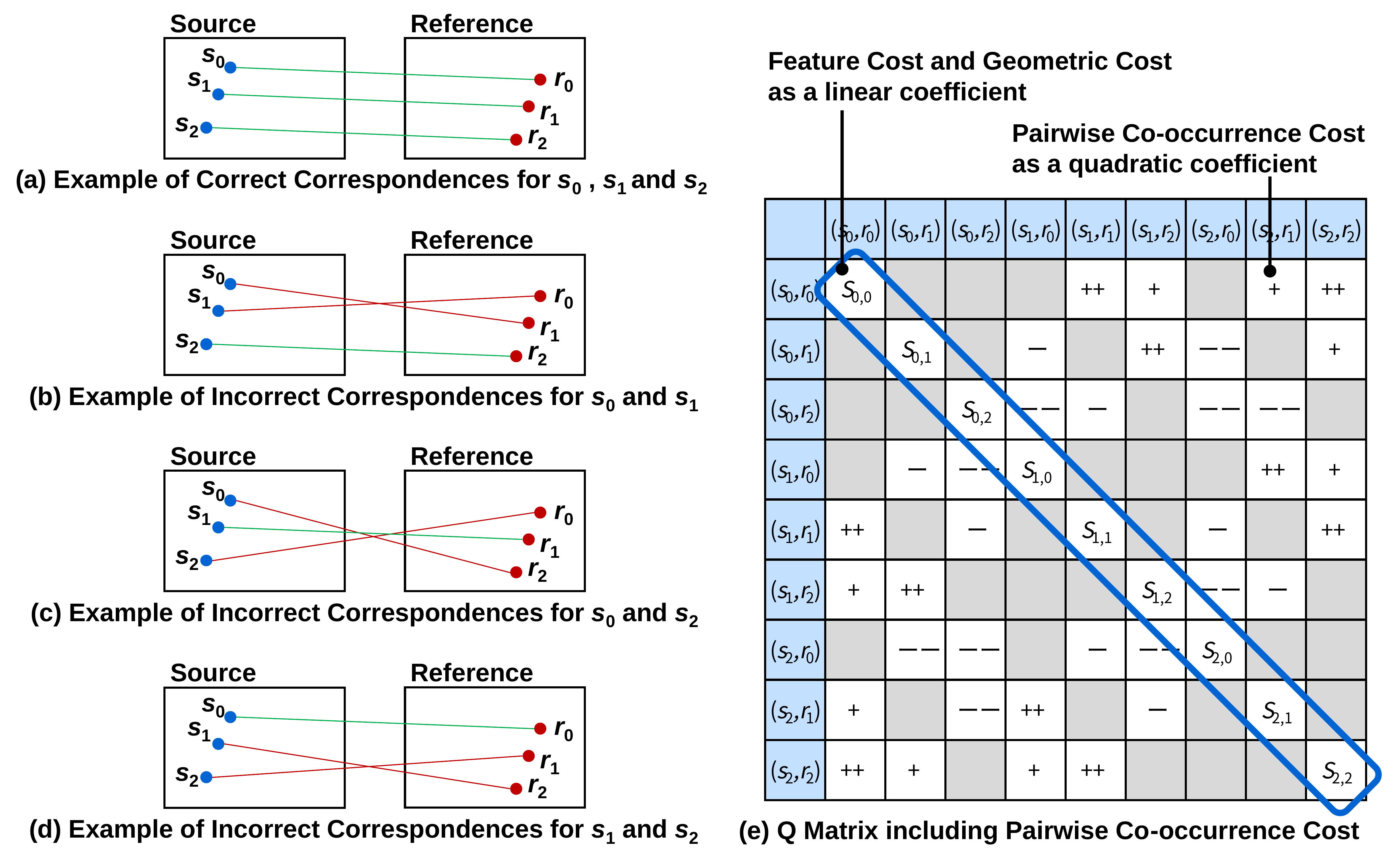}
\caption{Quadratic assignment problem using pairwise co-occurrence. (a) An example of correct matching. (b), (c), and (d) Examples of incorrect matching. (e) A tabular representation of a QUBO coefficient matrix (Q matrix) including feature, geometric, and pairwise co-occurrence costs.}
\label{fig:yamad4_cooccurrence}
\end{figure*}
\else
\Figure[t!](topskip=0pt, botskip=0pt, midskip=0pt)[scale=0.20]{figure/yamad4.pdf}
{ \textbf{Quadratic assignment problem using pairwise co-occurrence. (a) An example of correct matching. (b), (c), and (d) Examples of incorrect matching. (e) A tabular representation of a QUBO coefficient matrix (Q matrix) including feature, geometric, and pairwise co-occurrence costs.}
\label{fig:yamad4_cooccurrence} }
\fi

%% file: src/fig_yamad5.tex
\ifdefined\draft
\begin{figure*}[!ht]
\centering
\includegraphics[scale=0.20]{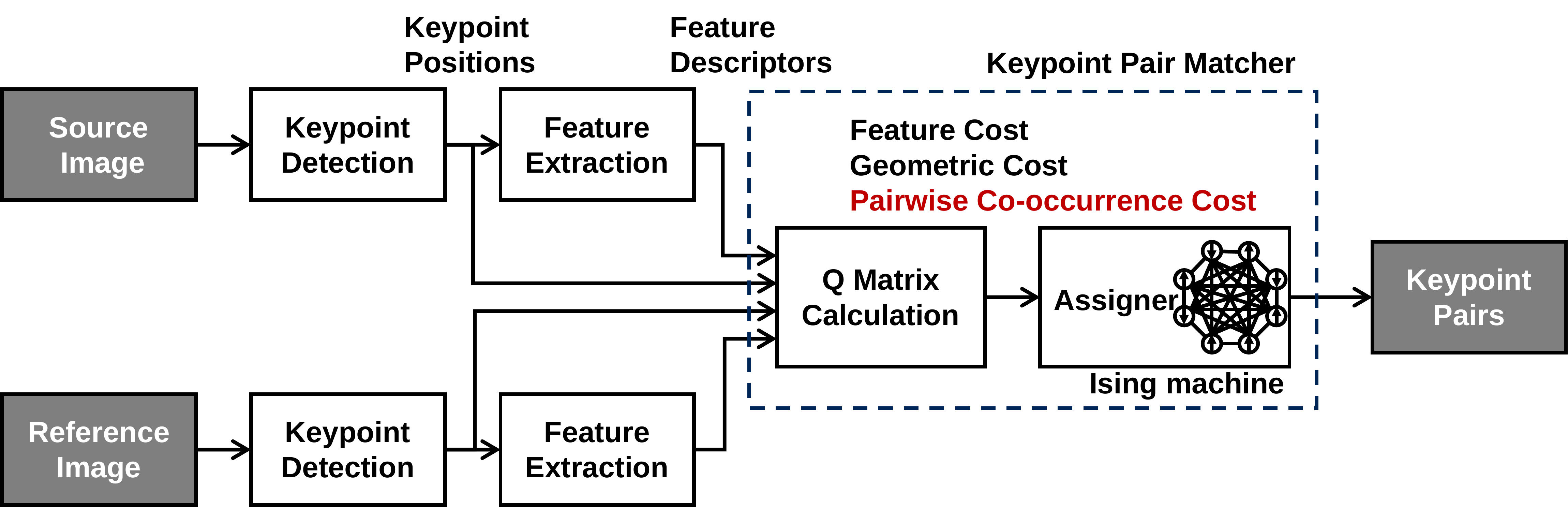}
\caption{Block diagram of local feature matching based on a co-occurrence-aware assignment problem.
}
\label{fig:yamad5_qubolfm}
\end{figure*}
\else
\Figure[t!](topskip=0pt, botskip=0pt, midskip=0pt)[scale=0.20]{figure/yamad5.pdf}
{ \textbf{Block diagram of local feature matching based on a co-occurrence-aware assignment problem.
}
\label{fig:yamad5_qubolfm}}
\fi

%% file: src/fig_yamad6.tex
\ifdefined\draft
\begin{figure*}[!ht]
\centering
\includegraphics[scale=0.20]{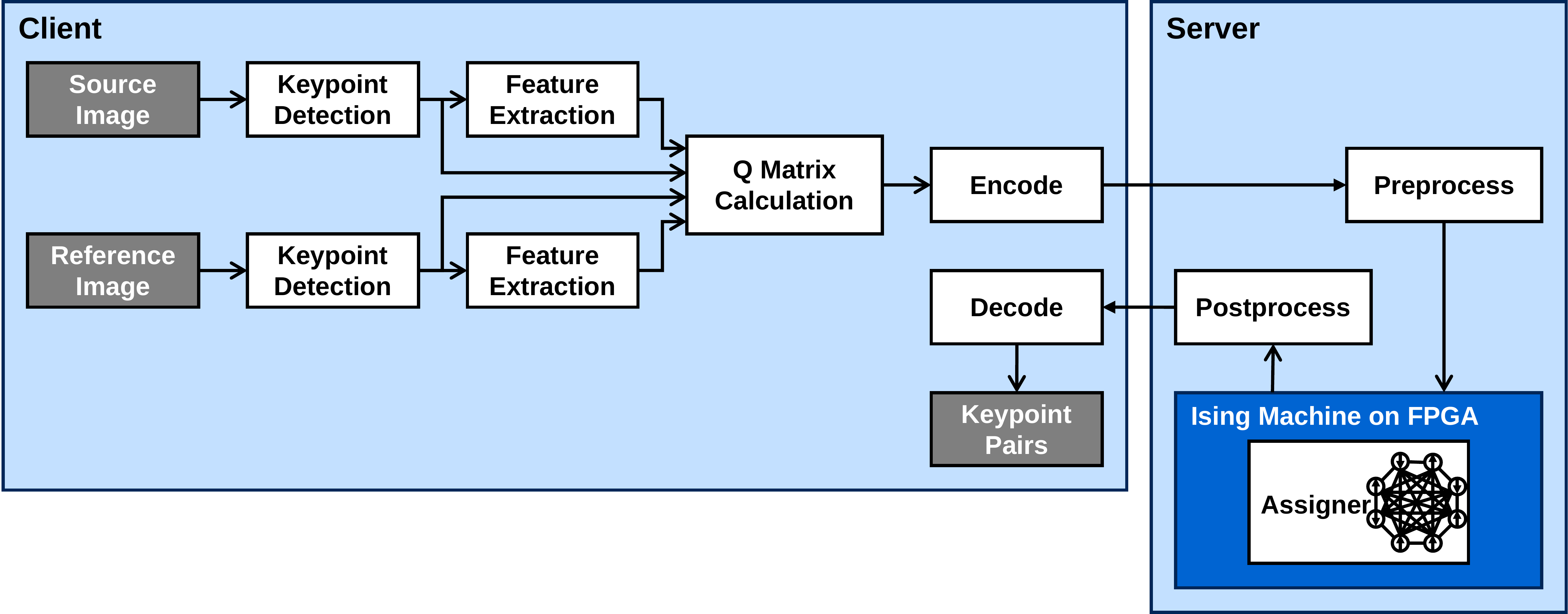}
\caption{System architecture for local feature matching, comprising a client and a server equipped with an FPGA-based Ising machine.}
\label{fig:yamad6_sysarch}
\end{figure*}
\else
\Figure[t!](topskip=0pt, botskip=0pt, midskip=0pt)[scale=0.20]{figure/yamad6.pdf}
{ \textbf{System architecture for local feature matching, comprising a client and a server equipped with an FPGA-based Ising machine.}
\label{fig:yamad6_sysarch}}
\fi

%% file: src/tab_yamad2.tex
\ifdefined\draft
\begin{table*}
\else
\begin{table}
\fi
\centering
\caption{\textbf{Quantitative evaluation for sample images in HPatches dataset \cite{hpatches_2017_cvpr}. Higher is better $(\uparrow)$, and lower is better $(\downarrow)$.}} \label{table}
\setlength{\tabcolsep}{3pt}
\begin{tabular}{|l|r|r|r|r|}
\toprule
Matcher & $N_{\rm{matches}}$ $(\uparrow)$ & TP $(\uparrow)$ & FP $(\downarrow)$ & Precision (\%) $(\uparrow)$ \\
\midrule
Baseline (NN)          & 185.30 & 108.36 & 76.94 & 52.53 \\
\midrule
Ising(F)               & 208.33 & 122.36 & 85.97 & 52.56 \\
Ising(F, C)            & 198.35 & 128.89 & 69.70 & 56.46 \\
Ising(F, G)            & 159.57 & 108.80 & 50.77 & 57.71 \\
Ising(F, G, C)         & 162.53 & 118.83 & 43.86 & 60.88 \\
\bottomrule
\end{tabular}
\label{tab:yamad2_hpatches}
\ifdefined\draft
\end{table*}
\else
\end{table}
\fi

%% file: src/fig_yamad7.tex
\ifdefined\draft
\begin{figure*}[!ht]
\centering
\includegraphics[width=7.0in]{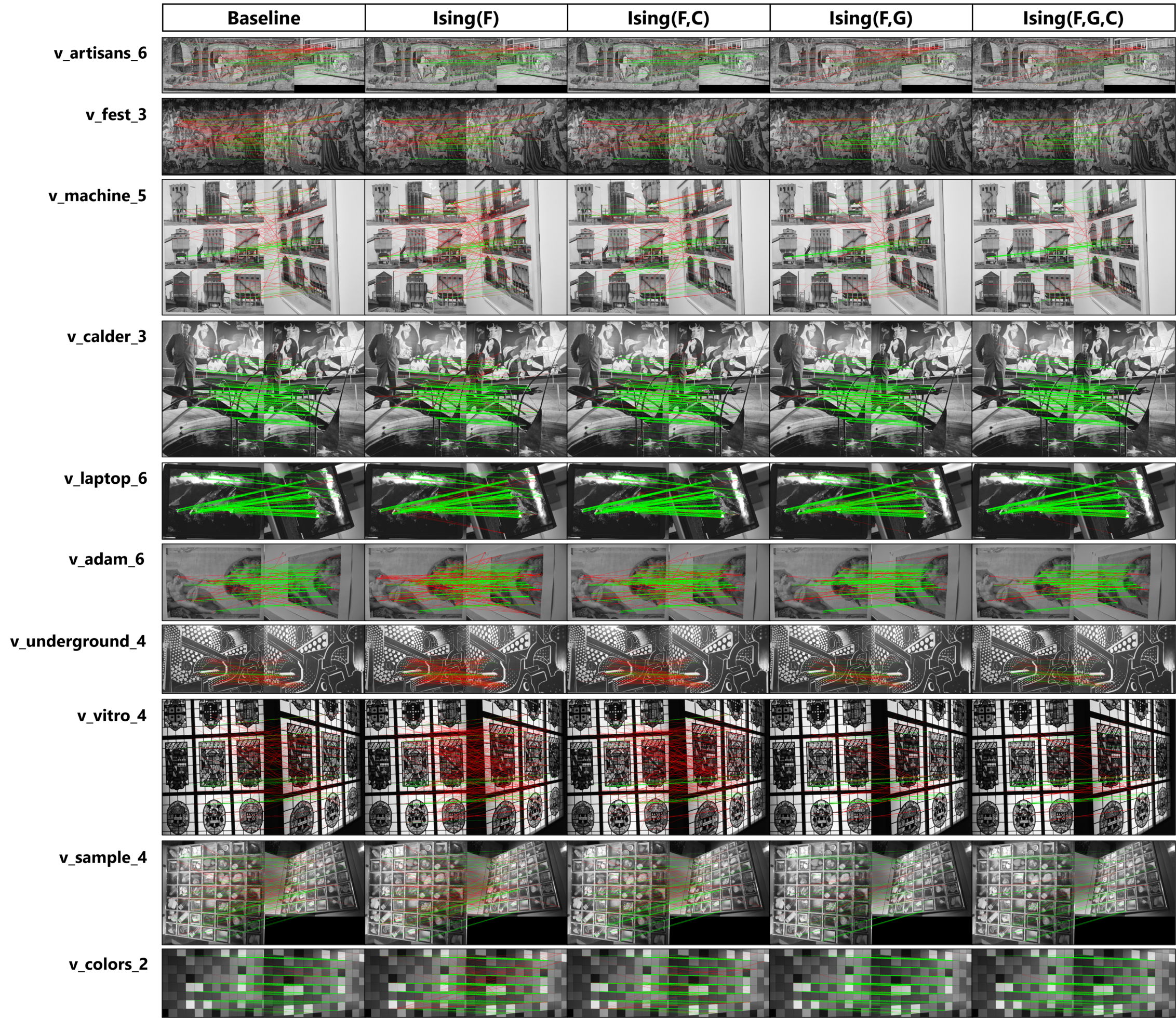}
\caption{Qualitative evaluation for sample images in HPatches dataset \cite{hpatches_2017_cvpr}. The green and red lines indicate correct and incorrect correspondences, respectively.}
\label{fig:yamad7_hpatches}
\end{figure*}
\else
\Figure[t!](topskip=0pt, botskip=0pt, midskip=0pt)[width=7.0in]{figure/yamad7.png}
{ \textbf{Qualitative evaluation for sample images in HPatches dataset \cite{hpatches_2017_cvpr}. The green and red lines indicate correct and incorrect correspondences, respectively.}
\label{fig:yamad7_hpatches}}
\fi

%% file: src/tab_yamad3.tex
\begin{table*}
\centering
\caption{\textbf{Quantitative evaluation of pose error and matching performance for the baseline (ORB-SLAM3 \cite{slam01-3_campos2021orb}) and the proposed method on driving scenes from the KITTI dataset \cite{Geiger2012CVPR}. Higher is better $(\uparrow)$, and lower is better $(\downarrow)$.}} \label{table:yamad3_kitti}
\setlength{\tabcolsep}{3pt}
\begin{tabular}{|l|r|r|r|r|r|r|r|}
\toprule
Seq    & Repetitive & \multicolumn{3}{|c|}{Baseline} & \multicolumn{3}{|c|}{Proposal} \\
       & Yes/No & APE (m) $(\downarrow)$ & RPE (m) $(\downarrow)$ & $N_{\rm{matches}}$ $(\uparrow)$ & APE (m) $(\downarrow)$ & RPE (m) $(\downarrow)$ & $N_{\rm{matches}}$ $(\uparrow)$ \\
\midrule
 03 &  No &   0.719 & 0.053 & 215.49 &   0.692 & 0.053 & 191.74 \\
 04 &  No &   1.142 & 0.078 & 165.16 &   0.987 & 0.075 & 140.78 \\
 08 &  No & 104.168 & 0.942 & 154.74 & 108.808 & 1.195 & 124.93 \\
 10 &  No &  14.794 & 0.226 & 175.98 &  17.019 & 0.308 & 133.93 \\
\midrule
 01 & Yes & 521.161 & 7.377 & 152.92 & 137.728 & 2.589 & 135.27 \\
\bottomrule
\end{tabular}
\label{tab:yamad3_kitti}
\end{table*}

%% file: src/fig_yamad8.tex
\ifdefined\draft
\begin{figure}[!ht]
\centering
\includegraphics[scale=0.15]{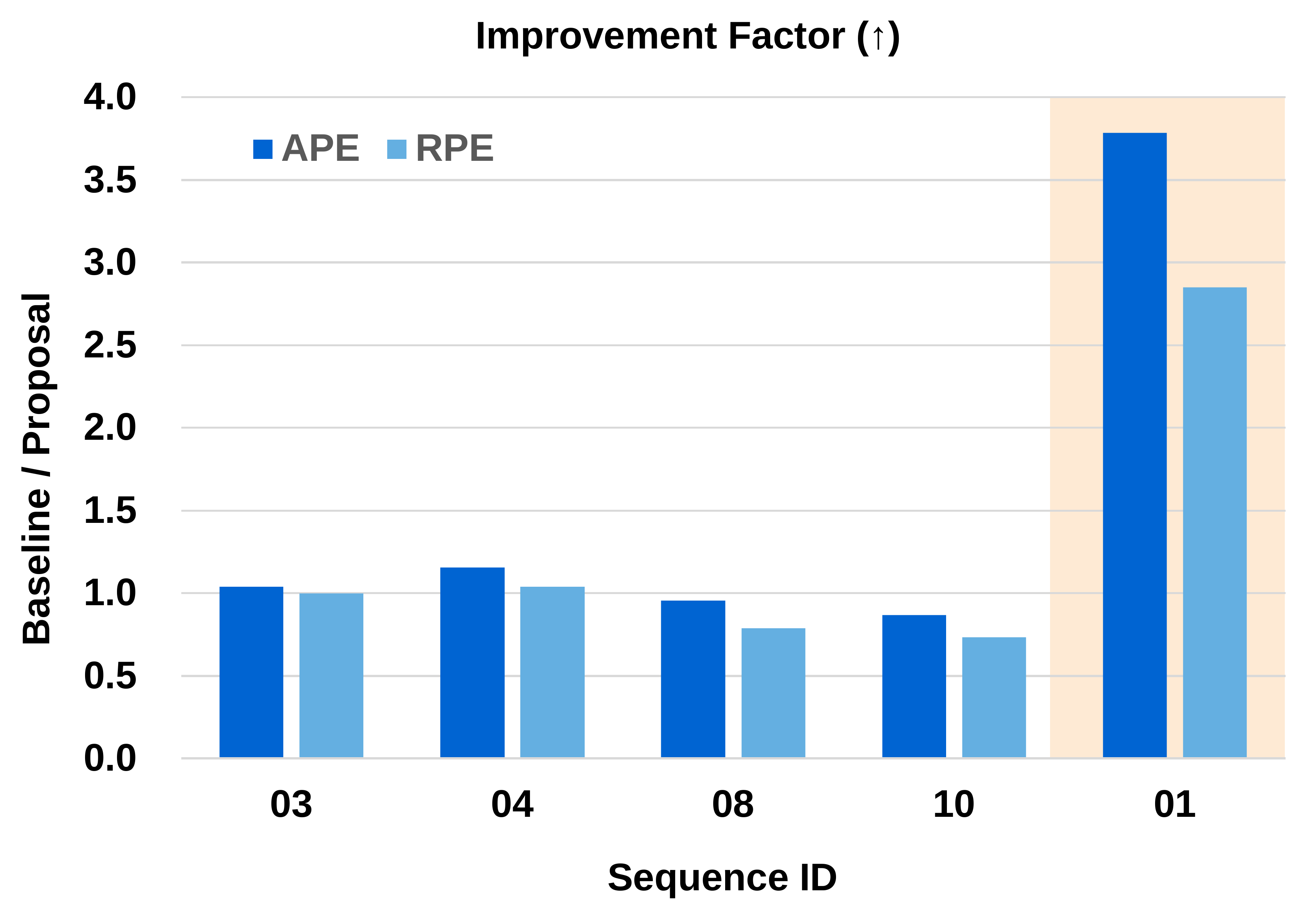}
\caption{Improvement factor of pose error achieved by the proposed method relative to the baseline (ORB-SLAM3 \cite{slam01-3_campos2021orb}) on the KITTI driving sequences \cite{Geiger2012CVPR}. Higher is better $(\uparrow)$.}
\label{fig:yamad8_kittigraph}
\end{figure}
\else
\Figure[!t](topskip=0pt, botskip=0pt, midskip=0pt)[scale=0.15]{figure/yamad8.pdf}
{ \textbf{Improvement factor of pose error achieved by the proposed method relative to the baseline (ORB-SLAM3 \cite{slam01-3_campos2021orb}) on the KITTI driving sequences \cite{Geiger2012CVPR}. Higher is better $(\uparrow)$.}
\label{fig:yamad8_kittigraph}}
\fi

%% file: src/fig_yamad9.tex
\ifdefined\draft
\begin{figure*}[!ht]
\centering
\includegraphics[width=7.0in]{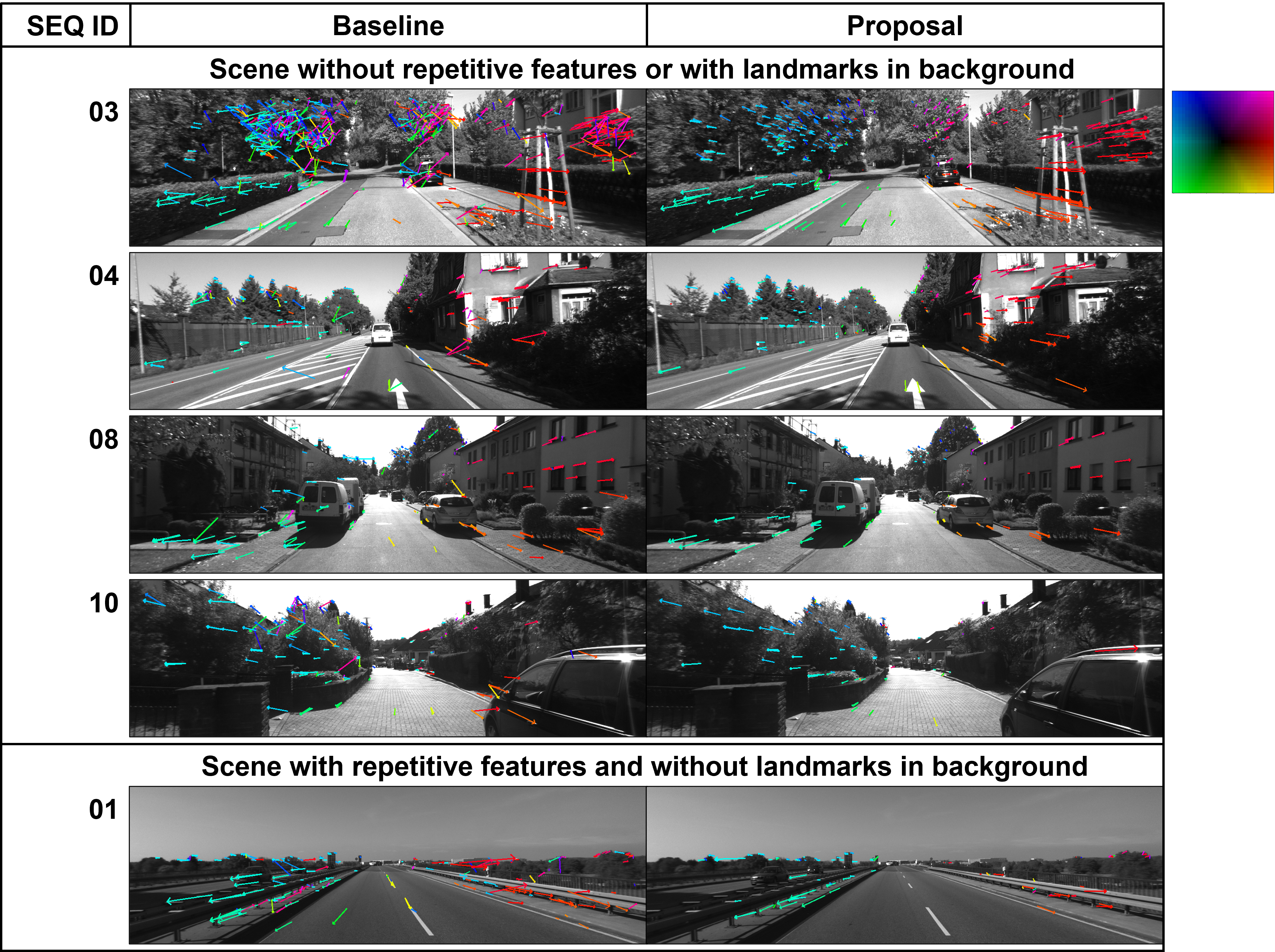}
\caption{Qualitative evaluation for sample images extracted from KITTI dataset \cite{Geiger2012CVPR}. The arrow colors indicate the motion direction and correspond to the color chart at the upper right.}
\label{fig:yamad9_kitti}
\end{figure*}
\else
\Figure[t!](topskip=0pt, botskip=0pt, midskip=0pt)[width=7.0in]{figure/yamad9.png}
{ \textbf{Qualitative evaluation for sample images extracted from KITTI dataset \cite{Geiger2012CVPR}. The arrow colors indicate the motion direction and correspond to the color chart at the upper right.}
\label{fig:yamad9_kitti}}
\fi

%% file: src/03_append_e.tex
% appendix
\ifdefined\draft
\else
\appendices
\fi

\section*{Appendices} \label{sec:appendix}

\subsection{QUBO and Ising problems} \label{sec:append-isingqubo}

The $N$-variable QUBO problem is to find a bit configuration (from among $2^{N}$ candidates) that minimizes the cost function:
\begin{equation}
H_{\text{QUBO}}(\mathbf{b}) = {\mathbf{b}}^{\text{T}} {\mathbf{Q}} {\mathbf{b}} = \sum_{i=1}^{N} \sum_{j=1}^{N} Q_{i,j} b_i b_j \label{eq:qubo}
\end{equation}
where $b_{i} \in \{0, 1\}$ denotes the $i$th bit, ${\mathbf{b}} = (b_{1}, \cdots , b_{N})$ is the vector representation of a bit configuration, $Q_{ij} (= Q_{ji})$ is a quadratic coefficient for the $i$th and $j$th bits, ${\mathbf{Q}}$ is the matrix representation of $\{ Q_{ij} \}$. Since $b_{i}^{2} = b_{i}$, the diagonal elements $Q_{ii}$ represent linear coefficients for $b_{i}$.

The $N$-variable Ising problem is to find a spin configuration that minimizes the Ising energy \cite{motcite45_barahona1982computational}:
\begin{equation}
H_{\text{Ising}}(\mathbf{s}) = -\frac{1}{2} {\mathbf{s}}^{\text{T}} {\mathbf{J}} {\mathbf{s}} + {\mathbf{h}}^{\text{T}} {\mathbf{s}} = - \frac{1}{2} \sum_{i=1}^{N} \sum_{j=1}^{N} J_{i,j} s_i s_j +  \sum_{i=1}^{N} h_i s_i \label{eq:ising}
\end{equation}
where \(s_i \in \{-1, +1\}\) is the $i$th Ising spin, ${\mathbf{s}} = (s_{1}, \cdots , s_{N})$ is the vector representation of a spin configuration, \(J_{i,j} (= J_{j,i}) \) is the coupling coefficient between the $i$th and $j$th spins $(J_{i,i} = 0)$, ${\mathbf{J}}$ is the matrix representation of $\{ J_{i,j} \}$, and \(h_i\) is a bias (or linear) coefficient for the $i$th spin, ${\mathbf{h}}$ is the vector representation of $\{ h_{i} \}$.

The QUBO problem in the form of Eq. (\ref{eq:qubo}) can be written as an equivalent Ising problem in the form of Eq. (\ref{eq:ising}) using the following relations:
\begin{align}
s_i &= 2b_i - 1 \\
J_{i,j} &=
\begin{cases}
- \frac{Q_{i,j}}{2} & {\text{if }}\ i \ne j \\
0                   & {\text{if }}\ i = j
\end{cases} \\
h_i &= \sum_{j=1}^{N} \frac{Q_{i,j}}{2}
\end{align}

\subsection{Simulated Bifurcation} \label{sec:append-sbm}

Simulated bifurcation (SB) \cite{motcite18_goto2019combinatorial, motcite20_goto2021high} is a quantum-inspired \cite{motcite18_goto2019combinatorial}, highly-parallelizable \cite{motcite19_tatsumura2019fpga, motcite21_tatsumura2021scaling, motcite23_kashimata2024efficient}, metaheuristic algorithm for computationally-hard combinatorial (or discrete) optimization.
SB-based Ising machine belongs to a group of oscillator-based Ising machines \cite{
motcite27_honjo2021100, motcite28_kalinin2020polaritonic, motcite29_bohm2019poor,
motcite34_graber2024integrated, motcite35_moy20221, motcite36_albertsson2021ultrafast, motcite37_wang2021solving,
motcite42_leleu2021scaling}. 
SB finds an optimal (exact) or near-optimal solution to the Ising problem by simulating the time-evolution process of coupled nonlinear oscillators according to Hamilton's equations of motion (without energy-dissipative or noise-based mechanisms).
The SB has several variants including adiabatic SB, ballistic SB, and discrete SB, which differ in terms of nonlinearity \cite{motcite67_bohm2021order} and discreteness \cite{motcite20_goto2021high}.
In the SB, the $i$th nonlinear oscillator corresponds to the $i$th Ising spin and its state is described by the position and momentum $(x_i, y_i)$. 
The update procedure of $x_i$ and $y_i$ for the ballistic SB, used in this work, is as follows \cite{motcite20_goto2021high}.
\begin{align}
y_i^{t_{k+1}} &\leftarrow y_i^{t_{k}} + [-(a_0 - a^{t_{k}})x_i^{t_k} - \eta h_i + c_0 \sum_{j}^{N} J_{i,j} x_j^{t_k}] \Delta_t, \\
x_i^{t_{k+1}} &\leftarrow x_i^{t_{k}} + a_0 y_i^{t_{k+1}} \Delta_t, \\
(x_i^{t_{k+1}}, y_i^{t_{k+1}}) &\leftarrow 
\begin{cases}
        (\text{sgn}(x_i^{t_{k+1}}), 0) & \text{if } |x_i^{t_{k+1}}| >    1, \\
(x_i^{t_{k+1}}, y_i^{t_{k+1}}) & \text{if } |x_i^{t_{k+1}}| \leq 1,
\end{cases} \label{eq:sbm_xiyi}
\end{align}
where \(a_0\), \(c_0\) and \(\eta\) are positive constants, \(a^{t_k}\) is a control parameter increasing from zero to \(a_0\), and \(\text{sgn}(x)(= \pm 1)\) is the sign function.
Eq. \ref{eq:sbm_xiyi} is a nonlinear transfer function \cite{motcite67_bohm2021order}, physically corresponding to a perfectly inelastic wall existing at \(x = \pm 1\).
The time increment is denoted as \(\Delta_t\), and thus, \(t_{k+1} = t_k + \Delta_t\).
After iterating the update procedure for the predetermined time steps (\(N_{step}\)), the $i$th position \(x_i\) is digitized to be the \(i\)th spin (\(\pm 1\)) by taking the sign of \(x_i\).
In this work, \(a_0 = 1\), \(c_0 = \eta = 0.2\), \(\Delta_t = 0.2\), and \(N_{step} = 1000\).

\subsection{Parameter Configuration} \label{sec:append-param}

The coefficients shown in Eq. (\ref{eq:h-total}) were adjusted by varying the values of ${\alpha}_f$, ${\alpha}_g$, ${\alpha}_c$, ${\alpha}_p$, ${\beta}_f$, and ${\beta}_g$ based on the results of visual evaluations to determine appropriate values.
We confirmed that similar results were obtained for values near those shown in Section \ref{sec:eval-methodology}.

\section*{Description of Supplementary Files} \label{sec:supplementary}

\begin{itemize}
\item Supplementary Data 1: Movie (baseline vs. proposal) for KITTI Dataset Sequence 01
\item Supplementary Data 2: Movie (baseline vs. proposal) for KITTI Dataset Sequence 03
\item Supplementary Data 3: Movie (baseline vs. proposal) for KITTI Dataset Sequence 04
\item Supplementary Data 4: Movie (baseline vs. proposal) for KITTI Dataset Sequence 08
\item Supplementary Data 5: Movie (baseline vs. proposal) for KITTI Dataset Sequence 10
\end{itemize}
For each movie, the top side shows the result of the baseline, and the bottom side shows the result of the proposed method.

\section*{Acknowledgment}

The authors thank Satoru Jimbo for fruitful discussions and technical support.

\section*{Conflict of Interest}

K.T. is included in inventors on a US patent application related to this work filed by the Toshiba Corporation (US Patent Application No. 2021-0182356). Y.Y., Y.H., Y.I., and K.T. are included in inventors on a Japanese patent application related to this work filed by the Toshiba Corporation (Japanese Patent Application No. 2025-081016). The authors declare that they have no other conflict of interest.

\bibliography{src/refer}
\bibliographystyle{IEEEtranDOI}

\ifdefined\draft
\begin{figure}[ht]
\vspace{0.5cm}
\noindent\includegraphics[width=1in,height=1.25in,clip,keepaspectratio]{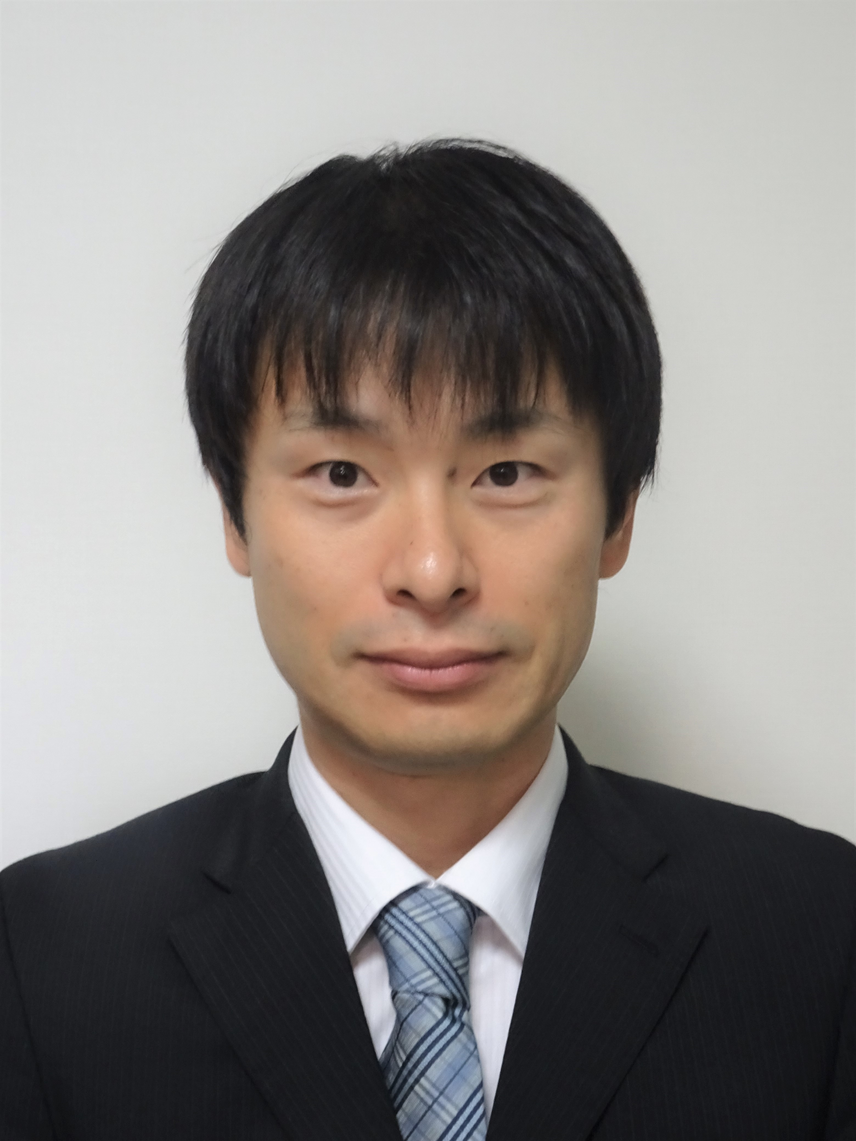}\\
{\small Yutaka Yamada
\else
\begin{IEEEbiography}
[{\includegraphics[width=1in,height=1.25in,clip,keepaspectratio]{./figure/Photo_YutakaYamada.png}}]
{Yutaka Yamada}
\fi
(Member, IEEE) received the B.E. and M.E. degrees in Information and Computer Science from Keio University, Yokohama, Japan, in 2003 and 2005, respectively.
He joined Toshiba Corporation, Kawasaki, Japan, in 2005.
He has been involved in research and development of reconfigurable processors, image recognition accelerators, image signal processors, and system architectures for consumer electronics and autonomous driving.
His current research interests include domain-specific computing and its hardware architecture, with a particular focus on quantum computing, neural networks, and signal processing for embedded systems.
\ifdefined\draft
}
\vspace{-0.5cm} 
\end{figure}
\else
\end{IEEEbiography}
\fi

\ifdefined\draft
\begin{figure}[ht]
\vspace{0.5cm}
\noindent\includegraphics[width=1in,height=1.25in,clip,keepaspectratio]{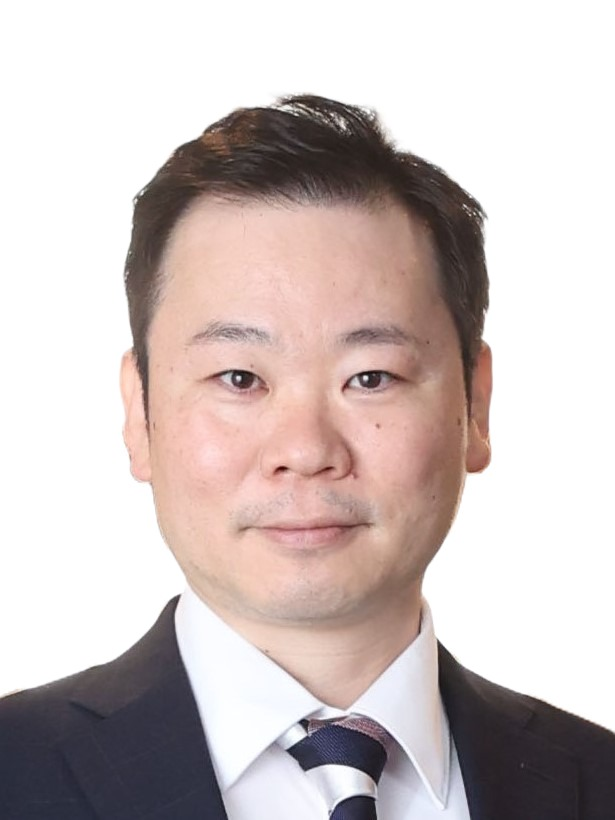}\\
{\small Yohei Hamakawa
\else
\begin{IEEEbiography}
[{\includegraphics[width=1in,height=1.25in,clip,keepaspectratio]{./figure/Photo_YoheiHamakawa.png}}]
{Yohei Hamakawa}
\fi
received the B.E. and M.E. degrees in Advanced Electronics and Optical Science from Osaka University, Japan, in 2000 and 2002, respectively.
He joined Toshiba Corporation in 2002. He was engaged in the development of image-processing engines for digital televisions and TV products, and distributed computing algorithms for deep learning.
His research interests include domain-specific computing, quantum computation, optimization in quantum circuit design, and their applications.
\ifdefined\draft
}
\vspace{-0.5cm} 
\end{figure}
\else
\end{IEEEbiography}
\fi

\ifdefined\draft
\begin{figure}[ht]
\vspace{0.5cm}
\noindent\includegraphics[width=1in,height=1.25in,clip,keepaspectratio]{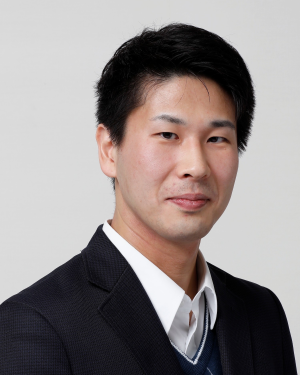}\\
{\small Yutaro Ishigaki
\else
\begin{IEEEbiography}
[{\includegraphics[width=1in,height=1.25in,clip,keepaspectratio]{./figure/Photo_YutaroIshigaki.png}}]
{Yutaro Ishigaki}
\fi
(Member, IEEE) graduated from Tokyo National College of Technology, Japan, in 2012. He received the B.S. and M.S. degrees in electrical and electronic engineering from Tokyo University of Agriculture and Technology, Japan, in 2014 and 2016, respectively.
In 2016, he joined Toshiba Corporation, Japan, where he was involved in the development of image recognition processors and hardware acceleration technologies for embedded localization systems.
His current research focuses on quantum cryptography and applications of quantum-inspired computing.
His research interests include domain-specific computing, hardware acceleration, and quantum communications.
\ifdefined\draft
}
\vspace{-0.5cm} 
\end{figure}
\else
\end{IEEEbiography}
\fi

\ifdefined\draft
\begin{figure}[ht]
\vspace{0.5cm}
\noindent\includegraphics[width=1in,height=1.25in,clip,keepaspectratio]{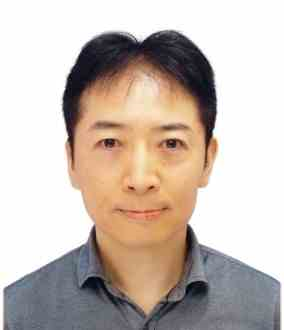}\\
{\small Masaya Yamasaki
\else
\begin{IEEEbiography}
[{\includegraphics[width=1in,height=1.25in,clip,keepaspectratio]{./figure/Photo_MasayaYamasaki.png}}]
{Masaya Yamasaki}
\fi
received the B.E. and M.E. degrees in computer science and communication engineering from Kyushu University, Japan, in 1997 and 1999, respectively.
He joined Toshiba Corporation, in 1999.
He was engaged in the development of image processing engines (interframe interpolation technology) for digital televisions (including ones with Cell Broadband Engines), FPGA-based coprocessors for multi-channel video recording and three-dimensional display, and industrial systems such as high-speed financial trading systems and autonomous control units. 
His research interests include domain-specific computing, high-level synthesis design space exploration, and proof-of-concept study with FPGA devices.
\ifdefined\draft
}
\vspace{-0.5cm} 
\end{figure}
\else
\end{IEEEbiography}
\fi

\ifdefined\draft
\begin{figure}[ht]
\vspace{0.5cm}
\noindent\includegraphics[width=1in,height=1.25in,clip,keepaspectratio]{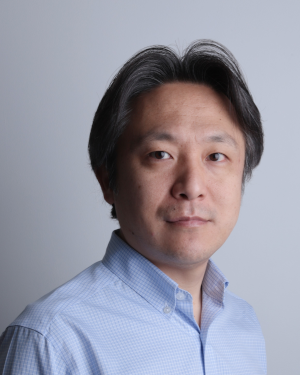}\\
{\small Kosuke Tatsumura
\else
\begin{IEEEbiography}
[{\includegraphics[width=1in,height=1.25in,clip,keepaspectratio]{./figure/Photo_KosukeTatsumura.png}}]
{Kosuke Tatsumura}
\fi
(Member, IEEE) received the B.E., M.E., and Ph.D. degrees in electronics, information and communications engineering from Waseda University, Japan, in 2000, 2001, and 2004, respectively. After working as a Postdoctoral Fellow with Waseda University, he joined the Corporate Research and Development Center (currently, the Corporate Laboratory), Toshiba Corporation, in 2006. There, he is currently a Senior Fellow, leading a research team and several projects toward realizing innovative industrial systems based on cutting-edge computing technology. He has been a Lecturer with Waseda University, since 2013. He was a Visiting Researcher with the University of Toronto, from 2015 to 2016. He received the Best Paper Award at IEEE International Conference on Field-Programmable Technology (FPT), in 2016, and the Awards for Science and Technology, the Commendation for Science and Technology by the Minister of Education, Culture, Sports, Science and Technology, Japan, in 2026.
\ifdefined\draft
}
\vspace{-0.5cm} 
\end{figure}
\else
\end{IEEEbiography}
\fi